\documentclass[aps,prd,preprint,eqsecnum,tightenlines,floats,
showpacs,superscriptaddress]{revtex4}
\usepackage{amssymb} 
\usepackage{hyperref} 
\usepackage{graphicx}
\usepackage{subfigure}

\begin{document}


\date{July 10, 2013}

\title{Axions and the Galactic Angular Momentum Distribution}

\author{N. Banik and P. Sikivie}
\affiliation{Department of Physics, University of Florida, 
Gainesville, FL 32611, USA}

\begin{abstract}

We analyze the behavior of axion dark matter before it falls into 
a galactic gravitational potential well.  The axions thermalize 
sufficiently fast by gravitational self-interactions that almost 
all go to their lowest energy state consistent with the total 
angular momentum acquired from tidal torquing.  That state is 
a state of rigid rotation on the turnaround sphere.  It predicts 
the occurrence and detailed properties of the caustic rings of dark 
matter for which observational evidence had been found earlier.  
We show that the vortices in the axion Bose-Einstein condensate 
(BEC) are attractive, unlike those in superfluid $^4$He and dilute 
gases.  We expect that a large fraction of the vortices in the 
axion BEC join into a single big vortex along the rotation axis 
of the galaxy. The resulting enhancement of caustic rings explains 
the typical size of the rises in the Milky Way rotation curve 
attributed to caustic rings.  We show that baryons and ordinary 
cold dark matter particles are entrained by the axion BEC and 
acquire the same velocity distribution.  The resulting baryonic 
angular momentum distribution gives a good qualitative fit to the 
distributions observed in dwarf galaxies.  We give estimates of 
the minimum fraction of dark matter that is axions.

\end{abstract}

\maketitle

\section{Introduction}

One of the outstanding problems in science today is the identity of the 
dark matter of the universe \cite{PDM}.  The existence of dark matter is
implied by a large number of observations, including the dynamics of galaxy  
clusters, the rotation curves of individual galaxies, the abundances of light
elements, gravitational lensing, and the anisotropies of the cosmic microwave
background radiation.  The energy density fraction of the universe in dark
matter is observed to be 26.7\% \cite{Planck}.  The dark matter must be
non-baryonic, cold and collisionless.  {\it Non-baryonic} means that the 
dark matter is not made of ordinary atoms and molecules.  {\it Cold} means 
that the primordial velocity dispersion of the dark matter particles is 
sufficiently small, less than about
$10^{-8}~c$ today, so that it may be set equal to zero as far as the formation
of large scale structure and galactic halos is concerned.  {\it Collisionless}
means that the dark matter particles have, in first approximation, only
gravitational interactions.  Particles with the required properties are   
referred to as `cold dark matter' (CDM).  The leading CDM candidates are
weakly interacting massive particles (WIMPs) with mass in the 100 GeV range, 
axions with mass in the $10^{-5}$ eV range, and sterile neutrinos with mass  
in the keV range.  To  try and tell these candidates apart on the basis of   
observation is a tantalizing quest.

Recently it has been argued that the dark matter is axions \cite{CABEC,
case,therm} because this assumption provides a natural explanation and 
detailed account of the existence and properties of caustic rings of 
dark matter in galactic halos.  Axions are different from the other 
cold dark matter candidates, such as WIMPs and sterile neutrinos, because 
axions form a Bose-Einstein condensate (BEC).  They do so as a result of 
their gravitational self-interactions.  On time scales long compared to 
their thermalization time scale $\tau$, almost all axions go to the 
lowest energy state available to them.  The other dark matter candidates
do not do this.  The (re)thermalization of the axion BEC is sufficiently 
fast that axions that are about to fall into a galactic gravitational 
potential well go to their lowest energy available state consistent with 
the total angular momentum they acquired from nearby protogalaxies through 
tidal torquing \cite{therm}.  That state is a state of net overall rotation.  
In contrast, ordinary cold dark matter falls into galactic gravitational 
potential wells with an irrotational velocity field \cite{inner}. The inner 
caustics are different in the two cases.  In the case of net overall rotation, 
the inner caustics are rings \cite{crdm} whose cross-section is a section of 
the elliptic umbilic $D_{-4}$ catastrophe \cite{sing}, called caustic rings 
for short.  If the velocity field of the infalling particles is irrotational, 
the inner caustics have a `tent-like' structure which is described in detail in 
ref.~{\cite{inner} and which is quite distinct from caustic rings.  Evidence was 
found for caustic rings.  A summary of the evidence is given in ref.~\cite{MWhalo}.  
Furthermore, it was shown in ref.~\cite{case} that the assumption that the dark 
matter is axions explains not only the existence of caustic rings but also their 
detailed properties, in particular the pattern of caustic ring radii and their 
overall size.  

The main purpose of the present paper is to give a more detailed description 
of the behavior of axion dark matter before it falls into the gravitational 
potential well of a galaxy.  In particular we want to investigate the 
appearance and evolution of vortices in the rotating axion BEC, and ask 
whether they have implications for observation.  We obtain two results 
that may be somewhat surprising.  The first is that, unlike the vortices 
in superfluid $^4$He and in BECs of dilute gases, the vortices in axion 
BEC attract each other.  The reason for the difference in behavior is that 
atoms have short range repulsive interactions whereas axions do not.  The 
vortices in the axion BEC join each other producing vortices of ever increasing 
size.  When two vortices join, their radii (not their cross-sectional areas) are 
added.  We expect a huge vortex to form along the rotation axis of the galaxy as 
the outcome of the joining of numerous smaller vortices.  We call it the `big
vortex'.  The presence of a big vortex implies that the infall is not isotropic 
as has been  assumed in the past \cite{STW,crdm,sing,MWhalo}.  The axions fall 
in preferentially along the equatorial plane.  Caustic rings are enhanced as  
a result because the density of the flows that produce the caustic rings is 
larger. We propose this as explanation of the fact that the rises in the Milky 
Way rotation curve attributed to caustic rings \cite{MWcr} are typically a 
factor five larger than predicted assuming that the infall is isotropic, a 
puzzle for which no compelling explanation had been given in the past.

The second perhaps surprising finding is that baryons are entrained 
by the axion BEC and acquire the same velocity distribution as the 
axion BEC.  The underlying reason for this is that the interactions
through which axions thermalize are gravitational and gravity is universal.  
The condition for the baryons to acquire net overall rotation by thermal 
contact with the axion BEC is the same as the condition for the axions 
to acquire net overall rotation by thermalizing among themselves.  When 
baryons and axions are in thermal equilibrium, their velocity fields 
are the same since otherwise entropy can be generated by energy-momentum 
exchanging interactions between them.  We expect that a big vortex forms 
in the baryon fluid as well, although one of lesser size than the big vortex 
in the axion fluid.  The resulting angular momentum distribution of baryons 
agrees qualitatively with that observed by van den Bosch et al. in dwarf 
galaxies \cite{Bosch}. In contrast, if the dark matter is WIMPs or sterile 
neutrinos, the predicted angular momentum distribution of baryons in galaxies 
differs markedly from the observed distribution, a discrepancy known as the 
`galactic angular momentum problem' \cite{Nava,Burk}.

We consider the possibility that the dark matter is a mixture of axions 
and another form of cold dark matter.  For the purposes of our discussion 
WIMPs and sterile neutrinos behave in the same way. So we call the other 
form of cold dark matter WIMPs for the sake of brevity.  There is a minimum 
fraction of dark matter that must be axions for the axions to rethermalize 
by gravitational interactions before they fall into galactic gravitational 
potential wells.  That fraction is of order 3\%.  If the axion fraction 
of dark matter is larger than of order 3\%, the axions acquire net overall 
rotation through their thermalizing gravitational self-interactions and 
entrain the baryons and WIMPs with them.  The baryons and WIMPs acquire 
the same velocity distribution as the axions before falling onto galactic 
halos.  The WIMPs therefore produce the same caustic rings, and at the same 
locations, as the axions.  However, to account for the typical size of the 
rises in the Milky Way rotation curve attributed to caustic rings, we find
that the fraction of dark matter in axions must be of order 37\% or more.

To investigate the issues of interest here we generalize the statistical 
mechanics  of many body systems in thermal equilibrium to the case when 
the system is rotating and total angular momentum is conserved.  When 
total angular momentum is conserved, a system of identical particles 
in thermal equilibrium is characterized by an angular frequency $\omega$ 
in addition to its temperature $T$ and its chemical potential $\mu$.  Broadly 
speaking, the state of thermal equilibrium of such systems is one of rigid 
rotation with angular frequency $\omega$.  Incidentally, we find that there 
is no satisfactory generalization of the isothermal sphere model of galactic 
halos with $\omega \neq 0$.  This is a serious flaw of that model since 
galactic halos acquire angular momentum from tidal torquing.  For a 
Bose-Einstein condensate, we derive the state that most particles condense 
into when $\omega \neq 0$.  In superfluid $^4$He this state is one of rigid 
rotation except for a regular array of vortices embedded in the fluid.  For 
a BEC of collisionless particles contained in a cylindrical volume the state 
is one of quasi-rigid rotation with all the particles as far removed from the 
axis of rotation as allowed by the Heisenberg uncertainty principle.  For an 
axion BEC about to fall into a galactic gravitational well the state is one 
in which each spherical shell rotates rigidly but the rotational frequency 
varies with the shell's radius $r$ as $r^{-2}$. (The motion is similar to 
that of water draining through a hole in a sink).  If the axions were to 
equilibrate fully, they would all move very close to the equatorial plane.  
Because their thermalization rate is not much larger than the Hubble rate, 
we expect that the axions start to move towards the equator but there is 
not enough time for the axions to get all localized there.  The motion of 
the axions toward the equatorial plane is another way to see why the big 
vortex forms. 

There is a large and growing literature exploring the hypothesis that 
the dark matter is a Bose-Einstein condensate of spin zero particles 
with mass of order $10^{-21}$ eV, or so \cite{dmBEC,RS}.  When the mass 
is that small, the de Broglie wavelength of the BEC in galactic halos 
is large enough (of order kpc) that the wave nature of the BEC has 
observable effects.  In contrast, our proposal is that the dark matter 
is composed of ordinary QCD axions, or of axion-like particles with 
properties similar to QCD axions. That QCD axions form a BEC is not 
an assumption on our part but a consequence of their standard properties.  
The axion BEC forms because the axions thermalize as a result of their 
gravitational interactions \cite{CABEC,therm}.  The axion BEC behaves 
differently from WIMPs because it rethermalizes on time scales less 
than the age of the universe.  The process of thermalization is key 
to understanding the properties of any BEC. It is only by thermalizing 
that a macroscopically large fraction of degenerate identical bosons go 
to their lowest energy available state.  The process of thermalization 
is not described by the Gross-Pitaevskii equation.  That equation 
describes the properties of the state that the particles condense 
into, but not the process by which the particles condense into that 
state.  We emphasize in Section II that vortices appear in a BEC 
only as part of the process of rethermalization.  The Gross-Pitaevskii 
equation describes the properties of vortices and their motions, but 
not their appearance.

The outline of this paper is as follows.  In Section II, we 
generalize the rules of statistical mechanics to the case where
the many body system conserves angular momentum.  We derive the 
rule that determines the state that a rotating BEC condenses into.
We verify that the rule is consistent with the known behaviour of 
superfluid $^4$He and derive the expected behavior of a rotating 
BEC of quasi-collisionless particles.  In Section III, we obtain 
the expected behavior of an axion BEC about to fall into a galactic 
gravitational potential well, and the response of baryons and 
WIMPS to the presence of the axion BEC.  In Section IV, we show 
that the axion BEC provides a solution to the galactic angular 
problem and derive a minimum fraction of dark matter in axions 
(37\%) from the typical size of rises in the Milky Way rotation 
curve attributed to caustic rings.  Section V provides a summary.

\section{Statistical mechanics of rotating systems}

In this section we discuss theoretical issues related to the 
statistical mechanics of rotating many-body systems.  First we 
generalize the well-known equilibrium Bose-Einstein, Fermi-Dirac 
and Maxwell-Boltzmann distributions to the case where angular 
momentum is conserved.  Systems which conserve angular momentum 
are characterized by an angular velocity, in addition to their 
temperature and chemical potential.  Next we discuss the  
self-gravitating isothermal sphere \cite{Chandra} as a model 
for galactic halos.  We show that the model is a reasonably 
good description only when the angular momentum is zero.  Next 
we discuss rotating Bose-Einstein condensates (BEC) and obtain 
the rule that determines the state which particles condense into.  
We analyze the properties of the vortices that must be present 
in any rotating Bose-Einstein condensate, and discuss the 
contrasting behaviours of vortices in superfluid $^4$He and 
in a fluid of quasi-collisionless particles.  Finally, we 
identify rethermalisation as the mechanism by which vortices 
appear in a BEC after it has been given angular momentum.

\subsection{Temperature, chemical potential and angular velocity}

A standard textbook result gives the average occupation number 
${\cal N}_i$ of particle state $i$ in a system composed of a 
huge number of identical particles at temperature $T$ and chemical 
potential $\mu$:
\begin{equation}
{\cal N}_i = {1 \over e^{{1 \over T}(\epsilon_i - \mu)} - \sigma}
\label{usdis}
\end{equation}
where $\epsilon_i$ is the energy of particle state $i$, and $\sigma$ 
= 0, +1 or -1.  If the particles are distinguishable, one must take 
$\sigma = 0$ and the distribution is called Maxwell-Boltzmann.  If 
the particles are bosons, $\sigma$ = + 1 and the distribution is 
called Bose-Einstein.  If the particles are fermions, $\sigma = -1$ 
and the distribution is called Fermi-Dirac.

To obtain Eq.~(\ref{usdis}), one considers a system with given 
total energy $E = \sum_i n_i \epsilon_i$ and given total number 
of particles $N = \sum_i n_i$.  The ${\cal N}_i$ are the values 
of the $n_i$ which maximize the entropy \cite{Huang,Pathria}. One 
may repeat this exercise in the case of a system that conserves total 
angular momentum $L = \sum_i n_i l_i$ about some axis, say $\hat{z}$.  
Maximizing the entropy for given total energy $E$, total number of 
particles $N$ and total angular momentum $L$, one finds: 
\begin{equation} 
{\cal N}_i
= {1 \over e^{{1 \over T}(\epsilon_i - \mu - \omega l_i)} - \sigma} 
\label{moddis}
\end{equation} 
where $\omega$ is an angular velocity.  The system at equilibrium 
is characterized by $T$, $\mu$ and $\omega$.  If the total number 
of particles is not conserved, one must set $\mu =0$.  Likewise if 
total angular momentum is not conserved one must set $\omega = 0$.

\subsection{The self-gravitating isothermal sphere revisited}

Consider a huge number of self-gravitating identical classical particles.  
(Although identical, they are disinguishable by arbitrarily unobtrusive 
labels.)  A particle state is given by its location $(\vec{r}, \vec{v})$ 
in phase-space.  According to Eq.~(\ref{moddis}), the particle density in
phase-space is given at thermal equilibrium by
\begin{equation}
{\cal N}(\vec{r}, \vec{v}) = {\cal N}_0 ~
e^{- {m \over T}[{1 \over 2} \vec{v}\cdot\vec{v} + \Phi(\vec{r})
- \omega \hat{z}\cdot(\vec{r} \times \vec{v})]}
\label{clgrav}
\end{equation}
where $m$ is the particle mass, ${\cal N}_0 \equiv e^{\mu \over T}$, 
and $\Phi(\vec{r})$ is the gravitational potential.  Newtonian gravity 
is assumed.  The gravitational potential satisfies the Poisson equation:
\begin{equation}
\nabla^2 \Phi(\vec{r}) = 4 \pi G m n(\vec{r})
\label{Poisson}
\end{equation} 
where 
\begin{equation}
n(\vec{r}) = \int d^3v~{\cal N}(\vec{r}, \vec{v})
\label{den}
\end{equation}
is the physical space density.

When $\omega = 0$, Eqs.~(\ref{clgrav}) and (\ref{den}) imply
\begin{equation}
n(\vec{r}) = n_0~e^{- {3 \over <v^2>} \Phi(\vec{r})}
\label{den2}
\end{equation}
where $<v^2> = {3 T \over m}$ is the velocity dispersion of the 
particles and $n_0 = {\cal N}_0 ({2 \pi <v^2> \over 3})^{3 \over 2}$.  
Combining Eqs.~(\ref{Poisson}) and (\ref{den2}) one obtains
\begin{equation}
\nabla^2 \Phi(\vec{r}) = 4 \pi G m n_0~e^{- {3 \over <v^2>} \Phi(\vec{r})}~~~\ .
\label{singeq}
\end{equation}
This equation permits a spherically symmetric ansatz, $\Phi(\vec{r}) = \Phi(r)$.
It can then be readily solved by numerical integration.  The solutions have the 
form
\begin{equation}
n(r) = n_0~d(r/s)
\label{sols}
\end{equation}
where
\begin{equation}
s = \left({<v^2> \over 12 \pi G m n_0}\right)^{1 \over 2}
\label{scale}
\end{equation} 
and $d(x)$ is a unique function with the limiting behaviours:
$d(x) \rightarrow 1$ as $x \rightarrow 0$ and $d(x) \rightarrow 2/x^2$
as $x \rightarrow \infty$.  A plot of the function $d(x)$ is shown, 
for example, in Fig. 1 of ref. \cite{galcen} or Fig. 4.7 of ref. \cite{BT}.  
The function $d(x)$ is often approximated by ${2 \over 2 + x^2}$ for 
convenience.  The phase-space distribution 
\begin{equation}
{\cal N}(\vec{r}, \vec{v}) = {\cal N}_0~
e^{ -{3 \over <v^2>} \vec{v}\cdot\vec{v}}~d(r/s)
\label{isofin}
\end{equation}
is called an `isothermal sphere' \cite{Chandra}.

The isothermal sphere is often used as a model for galactic halos 
\cite{SL,JKG}.  As such it has many attractive properties.  First, 
the isothermal sphere model is very predictive since it gives the full 
phase-space distribution in terms of just two parameters, $<v^2>$ and 
$s$.  Second, these two parameters are directly related to observable 
properties of a galaxy: $<v^2>$ is related to the galactic rotation 
velocity at large radii $v_{\rm rot}$ by 
$v_{\rm rot} = \sqrt{{2 \over 3} <v^2>}$ and $s$ is related to 
the galactic halo core radius $a$ by $a = \sqrt{2} s$.  Third, 
since $n(r) \propto 1/r^2$ for large $r$, the isothermal model 
predicts galactic rotation curves to be flat at large $r$. This 
is consistent with observation.  Fourth, since $n(r) \simeq n_0$ 
for small $r$, galactic halos have inner cores where the density 
is constant. This is also consistent with observation.  Fifth, the 
model is based on a simple physical principle, namely thermalization.

For all its virtues, we do not believe the isothermal model to be a good 
description of galactic halos.  The reason is that present day galactic 
halos, such as that of the Milky Way, are unlikely to be in thermal 
equilibrium.  If for some unexplained reason the Milky Way halo were 
in thermal equilibrium today, it would soon leave thermal equilibrium 
because it accretes surrounding dark matter.  The infalling dark matter 
particles only thermalize on time scales that are much longer than the 
age of the universe \cite{velpeak,rob}.  The flows of infalling dark 
matter produce peaks in the velocity distribution.  A large fraction 
of the halo, over 90\% in the model of ref. \cite{MWhalo}, is in cold 
flows.  This disagrees with the smooth Maxwell-Boltzmann distribution, 
Eq.~(\ref{isofin}), of the isothermal model.  The presence of 
infall flows, with high density contrast in phase-space, has 
been confirmed by cosmological N-body simulations \cite{DDT}.

Here we point to another flaw of the isothermal sphere as a model of 
galactic halos.  Galactic halos acquire angular momentum from tidal 
torquing.  If they are in thermal equilibrium, as the isothermal model 
supposes, the phase-space distribution must be given by Eq.~(\ref{clgrav}) 
with $\omega \neq 0$.  However, Eq.~(\ref{clgrav}) is an unacceptably 
poor description of galactic halos as soon as $\omega \neq 0$.  Indeed, 
Eq.~(\ref{clgrav}) may be rewritten
\begin{equation} 
{\cal N}(\vec{r}, \vec{v}) = {\cal N}_0 ~ 
e^{- {m \over T}[{1 \over 2} (\vec{v} - \vec{\omega}\times\vec{r})^2 
+ \Phi(\vec{r}) - {1 \over 2}(\vec{\omega}\times\vec{r})^2]} 
\label{clgrav2}
\end{equation} 
where $\vec{\omega} = \omega \hat{z}$.  Compared to the 
$\omega = 0$ case, the velocity distribution is locally boosted 
by the rigid rotation velocity $\vec{\omega}\times\vec{r}$.  The 
physical space density is 
\begin{equation} 
n(\vec{r}) = n_0 e^{- {m \over T}[\Phi(\vec{r}) -
{1 \over 2} (\vec{\omega}\times\vec{r})^2]}~~~\ .
\label{den3} 
\end{equation} 
Substituting this into Eq.~(\ref{Poisson}), one obtains 
\begin{equation}
\nabla^2 \Phi(\vec{r}) = 4 \pi G m n_0~
e^{{m \over T}[-  \Phi(\vec{r}) + {1 \over 2} \omega^2 \rho^2]}~~~\ .
\label{singeq2}
\end{equation}
where $(\rho, z, \phi)$ are cylindrical coordinates.  Eq.~(\ref{singeq2})
does not have any solutions for which the density $n(\vec{r})$ goes to 
zero for large $\rho$.  Indeed 
$d(\vec{r}) \propto e^{{m \over 2T} \omega^2 \rho^2}$ at large $\rho$
unless $\Phi \rightarrow {1 \over 2} \omega^2 \rho^2$ there.  But this 
implies, through Eq.~(\ref{Poisson}), that the density goes to the 
constant value ${\omega^2 \over 2 \pi G m}$ at large $\rho$. The 
particles at large $\rho$ have huge bulk motion with average velocity 
$\vec{\omega}\times \vec{r}$. Thus the rotating isothermal sphere is 
an object of infinite extent in a state of rigid rotation.  This is 
certainly inconsistent with the properties of galactic halos.

As was discussed by Lynden-Bell and Wood \cite{LB}, isothermal 
spheres and self-gravitating systems in general are unstable 
because their specific heat is negative, i.e. they get hotter 
when energy is extracted from them.  The instability implies 
a gravo-thermal catastrophe on a time scale which may be  
inconsistent with the age of galactic halos and thus cause 
further difficulties in using isothermal spheres as models
for galactic halos.  The instability of rotating isothermal 
spheres is discussed in ref. \cite{Chav}.

\subsection{The rotating Bose-Einstein condensate}

\subsubsection{Bose-Einstein condensation}

Bose-Einstein condensation occurs when the following four conditions 
are satisfied:  1) the system is composed of a huge number of identical 
bosons, 2) the bosons are highly degenerate, i.e. their average quantum 
state occupation number is larger than some critical value of order one, 
3) the number of bosons is conserved, and 4) the system is in thermal 
equilibrium.  When the four conditions are satisfied a fraction of order 
one of all the bosons are in the same state.  

Let us recall a simple argument \cite{Pethik} why Bose-Einstein
condensation occurs.  We first set $\omega = 0$.  For identical 
bosons, Eq.~(\ref{moddis}) states
\begin{equation}
{\cal N}_i = {1 \over e^{{1 \over T}(\epsilon_i - \mu)} - 1}~~\ .
\label{usdisb}
\end{equation}  
Let $i = 0$ be the ground state, i.e. the particle state with lowest
energy.  It is necessary that the chemical potential remain smaller than 
the ground state energy $\epsilon_0$ at all times since Eq.~(\ref{usdisb}) 
does not make sense for $\epsilon_i < \mu$.  The total number of particles 
$N(T, \mu) = \sum_i {\cal N}_i$ is an increasing function of $\mu$ for fixed 
$T$ since each ${\cal N}_i$ has that property.  Let us imagine that the total 
number $N$ of particles is increased while $T$ is held fixed.  The chemical 
potential increases till it reaches $\epsilon_0$.  At that point the total 
number of particles in excited ($i > 0$) states has its maximum value
\begin{equation}
N_{\rm ex}(T, \mu=\epsilon_0) = 
\sum_{i > 0} {1 \over e^{{1 \over T} (\epsilon_i - \epsilon_0)} - 1}~~\ .
\label{maxnum}
\end{equation}
In three spatial dimensions, $N_{\rm ex}(T, \mu=\epsilon_0)$ is finite 
\cite{Pethik}.  Consider what happens when, at a fixed temperature $T$, 
$N$ is made larger than $N_{\rm ex}(T, \mu =\epsilon_0)$.  The only possible 
system response is for the extra  $N - N_{\rm ex}(T, \mu =\epsilon_0)$ particles 
to go to the ground state.  Indeed the occupation number ${\cal N}_0$ of that 
state becomes arbitrarily large as $\mu$ approaches $\epsilon_0$ from below.  
This is the phenomenon of Bose-Einstein condensation.

We may repeat the above argument for $\omega \neq 0$.  In this 
case the chemical potential $\mu$ must remain smaller than the 
smallest $\eta_i \equiv \epsilon_i - \omega l_i$. The existence
of a minimum $\eta_i$ is guaranteed because the particle energy 
$\epsilon_i$ always contains a piece that is quadratic in $l_i$, 
namely the kinetic energy associated with motion in the $\hat{\phi}$ 
direction.  Let
\begin{equation}
\eta_0 \equiv \min_i [\epsilon_i - \omega l_i]
\label
{etao}
\end{equation}
and $i=0$ the particle state that minimizes $\eta$.  (We are
relabeling the states compared to the $\omega = 0$ case.)  The 
largest possible number of particles in states $i \neq 0$ at 
a given temperature $T$ is 
\begin{equation}
N_{\rm ex}(T, \mu=\eta_0) =
\sum_{i \neq 0} {1 \over e^{{1 \over T} (\eta_i - \eta_0)} - 1}~~\ .
\label{maxnum2}
\end{equation}
If the total number $N$ of particles is larger 
than $N_{\rm ex} (T, \mu = \eta_0)$ , the extra 
$N - N_{\rm ex} (T, \mu = \eta_0)$ particles go 
to the $i=0$ state.

\subsubsection{Fluid description of the condensed state}

Each particle state is described by a wavefunction $\Psi(\vec{r}, t)$
which satisfies the Schr\"odinger equation (we suppress for the time being
the index $i$ that labels particle states) 
\begin{equation} 
i \partial_t \Psi = 
[ - {1 \over 2 m} \nabla^2 + m \Phi(\vec{r}, t) + V(\vec{r}, t)] \Psi 
\label{Schro}
\end{equation} 
where $\Phi(\vec{r}, t)$ is the gravitational potential and $V(\vec{r}, t)$ 
is the non-gravitational potential energy of the particle in its background.  
We assume throughout that the particles are non-relativistic.  As is well-known, 
the probability density $\Psi^* \Psi$ and the probability flux density 
${1 \over m} Im(\Psi^* \vec{\nabla} \Psi)$ satisfy the continuity equation 
because the total probability to find the particle someplace is conserved.
 
When a state is occupied by a huge number $N$ of particles, that state's
wavefunction $\Psi(\vec{r}, t)$ describes the properties of a macroscopic
fluid.  The fluid density is 
\begin{equation} 
n(\vec{r}, t) = N \Psi^*\Psi 
\label{flden} 
\end{equation} 
and the fluid flux density is
\begin{equation} 
\vec{j}(\vec{r}, t) = N {1 \over m} Im(\Psi^*\vec{\nabla} \Psi)~~\ .  
\label{flflux} 
\end{equation} 
They satisfy the continuity equation 
\begin{equation} \partial_t n ~+~\vec{\nabla} \cdot \vec{j} = 0~~ 
\label{contin}
\end{equation} 
for the reason mentioned at the end of the previous paragraph.  The fluid 
velocity $\vec{v}(\vec{r},t)$ is defined by $\vec{v} \equiv {1 \over n} \vec{j}$.  
If we write the wavefunction as 
\begin{equation} \Psi(\vec{r}, t) \equiv B(\vec{r}, t)
e^{i \beta(\vec{r}, t)}~~\ , 
\label{Bbeta} 
\end{equation} 
then $n = N B^2$, $\vec{j} = N B^2 {1 \over m} \vec{\nabla} \beta$ and 
hence \cite{Pethik} 
\begin{equation} 
\vec{v}(\vec{r}, t) = {1 \over m} \vec{\nabla} \beta~~~\ . 
\label{flvel}
\end{equation} 
When a state is occupied by a huge number $N$ of particles, its wavefunction
$\Psi(\vec{r},t)$ satisfies Eq.~(\ref{Schro}) as does the wavefunction of
any other state.  However, the gravitational potential $\Phi$ and the
potential energy $V$ may have important contributions from the $N$
particles themselves, i.e. 
\begin{equation} \Phi(\vec{r}, t) = - G m \int d^3 r^\prime
~ {n(\vec{r}^\prime, t) \over |\vec{r} - \vec{r}^\prime|} + ... 
\label{selfgrav}
\end{equation} 
where the dots indicate other contributions to the gravitational 
potential, and likewise
\begin{equation} 
V(\vec{r}, t) = \int d^3r^\prime~ V_p(\vec{r} - \vec{r}^\prime) 
n(\vec{r}^\prime, t) + ... 
\label{selfint} 
\end{equation}
if the $N$ particles interact pairwise with forces derived from a
potential $V_p$.  Upon substituting Eqs.~(\ref{selfgrav}) and/or
(\ref{selfint}), Eq.~(\ref{Schro}) takes on a non-linear form.  This
non-linear form of the Schr\"odinger equation is called the
Gross-Pitaevskii equation.

Eq.~(\ref{Schro}) implies an Euler-type equation for the fluid
velocity \cite{Pethik}
\begin{equation}
\partial_t \vec{v} + (\vec{v}\cdot\vec{\nabla}) \vec{v} = 
- \vec{\nabla} q - \vec{\nabla} \Phi - {1 \over m} \vec{\nabla} V
\label{Euler}
\end{equation}
where 
\begin{equation}
q(\vec{r}, t) = - {1 \over 2 m^2} {\nabla^2 \sqrt{n} \over \sqrt{n}}~~\ .
\label{que}
\end{equation}
Except for the $- \vec{\nabla} q$ term, Eq.~(\ref{Euler}) is 
the Euler equation for a fluid of classical particles moving 
in the potentials $\Phi$ and $V$.  The $- \vec{\nabla} q$ term 
is a consequence of the Heisenberg uncertainty principle and 
accounts, for example, for the intrinsic tendency of a wavepacket 
to spread.  However, the $- \vec{\nabla} q$ term is irrelevant
on scales large compared to the de Broglie wavelength.  {\it On 
such large length scales the fluid described by the Schr\"odinger
equation is indistinguishable from a fluid of classical particles 
moving in the potentials $\Phi$ and $V$.}

\subsubsection{Vortices \cite{Ons,Feyn}}

It may appear at first that the requirement 
$\vec{\nabla}\times\vec{v} = 0$, implied by Eq.~(\ref{flvel}), 
disagrees with the principle just stated, since a fluid of 
particles may have a rotational velocity field whereas the 
wave description allows apparently only irrotational flow.  
However, that appearance is deceiving because Eq.~(\ref{flvel}) 
is valid only where $\Psi \neq 0$.  Indeed $\beta$ is not well 
defined where $\Psi = 0$. The following example is instructive 
\begin{equation}
\Psi(\rho, \phi, t) = A J_l(k \rho)~e^{i l \phi}~ 
e^{ - i {k^2 \over 2 m} t}
\label{example}
\end{equation}
where $J_l$ is the Bessel function of index $l$.  This wavefunction 
solves the Schr\"odinger equation with $\Phi = V = 0$.  It describes
an axially symmetric flow with energy ${k^2 \over 2 m}$ and $z$-component 
of angular momentum $l$ per particle.  The fluid is clearly rotating
since Eq.(\ref{flvel}) implies the velocity field 
\begin{equation}
\vec{v} = {l \over m \rho} \hat{\phi}~~\ . 
\label{exvel}
\end{equation}
The curl of that velocity field vanishes everywhere except on 
the $z$-axis: $\vec{\nabla}\times\vec{v} = 
{2 \pi l \over m} \hat{z} \delta^2(x,y)$ where $x$ and $y$ are 
Cartesian coordinates in the plane perpendicular to $\hat{z}$. 
$\Psi$ vanishes on the $z$-axis if $l \neq 0$.  Furthermore, since 
the Bessel function $J_l(s) \simeq 0$ for $s<<l$ \cite{Abram}, the 
density implied by Eq.~(\ref{example}) is tiny for $\rho$ much less 
than the classical turnaround radius $\rho_c = l/k$.  Thus, on length 
scales large compared to $k^{-1}$, the fluid motion described by the 
wavefunction (\ref{example}) is the same as the motion of a fluid 
composed of classical particles, in agreement with the principle 
stated at the end of the previous paragraph.

That the motion of a fluid of cold dark matter particles can be 
described by a wavefunction  was emphasised by Widrow and Kaiser
some twenty years ago \cite{Widrow}.  Ordinary (non BEC) cold dark 
matter particles have irrotational flow because their rotational 
modes have been suppressed by the expansion of the universe and, 
when density perturbations start to grow,  Eq.(\ref{Euler}) 
implies $\vec{\nabla} \times \vec{v} = 0$ at all times if 
$\vec{\nabla} \times \vec{v} = 0$ initially \cite{inner}.  
The previous paragraph emphasizes that a fluid of cold dark 
matter particles can be described by a wavefunction whether or 
not the velocity field is irrotational.  The only requirement is 
that the wavefunction vanish on a set of lines, called vortices, 
if the velocity field is rotational. 

Each vortex carries an integer number of units of angular momentum, 
$l$ units in the example of Eq.~(\ref{example}).  The following rule 
applies.  Let ${\cal C}[\Gamma]$ be the circulation of the velocity 
field along a closed path $\Gamma$ 
\begin{equation}
{\cal C}[\Gamma] \equiv \oint_\Gamma~d\vec{r}\cdot\vec{v}(\vec{r}, t)~~\ .
\label{circ}
\end{equation}
If the fluid is represented by a wave, Eq.~(\ref{flvel}) implies
\begin{equation}
{\cal C}[\Gamma] = {\Delta \beta \over m} 
\equiv {2 \pi \over m} l[\Gamma]
\label{flcirc}
\end{equation}
where $\Delta \beta$ is the change of the phase $\beta$ when 
going around $\Gamma$.  Since the wavefunction is single valued, 
$l[\Gamma] \equiv {1 \over 2 \pi} \Delta \beta$ is an integer.  
The surface enclosed by a closed path with non-zero circulation must 
be traversed by vortices whose units of angular momentum add up to
$l[\Gamma]$.  Another rule is that as long as the fluid is described 
by a wavefunction $\Psi(\vec{r}, t)$, vortices cannot appear spontaneously 
in the fluid.  They can only move about.  Indeed, consider an arbitrary 
closed path $\Gamma$.  The total number of vortices encircled by $\Gamma$, 
counting a vortex with angular momentum $l$ as $l$ vortices, is determined 
by the constraint of Eq.~(\ref{flcirc}).  The only way the RHS of that 
equation can change is by having the wavefunction $\Psi$ vanish somewhere 
on $\Gamma$ and letting a vortex cross that curve.  Finally, a third rule: 
vortices must follow the motion of the fluid.  This is a corollary of 
Kelvin's theorem which states that, in gradient flow - i.e. if the RHS 
of Euler's equation is a gradient, as is the case for Eq.~(\ref{Euler}) - 
the circulation of the velocity field along a closed path that moves with 
the fluid is constant in time.

\subsubsection{Superfluid $^4$He}

Let us see how the above considerations apply first to the case of superfluid 
$^4$He, and next to the case of a collisionless fluid.  Of course, we have 
nothing new to say about superfluid $^4$He, which we discuss merely to build 
confidence in the general approach described so far.   We found that, when 
Bose-Einstein condensation occurs, a macroscopically large number of particles, 
say $N$, condense into the state with lowest $\eta = \epsilon - \omega l$.   
We have, in terms of the quantities 
defined earlier,
\begin{eqnarray}
\epsilon &=& \int d^3 r~\Psi^* (- {1 \over 2 m} \nabla^2 + 
m \Phi + V) \Psi \nonumber\\
&=& {1 \over N} \int d^3r~n~ [ {m \over 2} \vec{v}\cdot\vec{v}
+ m (q + \Phi) + V]
\label{eps}
\end{eqnarray}
and 
\begin{eqnarray}
l &=& 
\int d^3r~\Psi^* \hat{z}\cdot(\vec{r}\times{1 \over i} \vec{\nabla})\Psi
= \int d^3r~\Psi^* {1 \over i} {\partial \over \partial \phi} \Psi 
\nonumber\\
&=& {1 \over N} \int d^3r~n~ m \hat{z}\cdot(\vec{r}\times\vec{v})~~~\ .
\label{ell}
\end{eqnarray}
He atoms have an interparticle potential $V_p$ that describes forces
that are strongly repulsive at short range and weakly attractive at 
long range.  In the liquid state, the average interatomic distance is 
of order the atom size.  Thus the density of superfluid $^4$He has an 
approximately constant value $n_0$. In that case
\begin{equation}
\epsilon \simeq {n_0 \over N} \int d^3r~{m \over 2} 
\vec{v}\cdot\vec{v}~~~~~,~~
~~~l \simeq {n_0 \over N} \int d^3r ~m
\hat{z}\cdot(\vec{r}\times\vec{v})
\label{He4}
\end{equation}
and therefore
\begin{equation}
\eta \simeq {n_0 m \over 2 N} \int d^3r~[(v_z)^2 + (v_\rho)^2 +
(v_\phi - \omega \rho)^2 - \omega^2 \rho^2]~~~\ .
\label{Heta}
\end{equation}
$\eta$ is minimized when $\vec{v} = \omega \hat{z}\times\vec{r}$.
So, when superfluid $^4$He carries angular momentum, it is in a 
state of rigid rotation when viewed on length scales large compared 
to the de Broglie wavelength.  Because rigid rotation implies that 
the velocity field has non-zero curl, vortices are present. The 
vortices are parallel to the $z$-axis and their number density 
per unit area is 
\begin{equation}
{m \over 2 \pi} \hat{z}\cdot(\vec{\nabla}\times\vec{v}) = 
{m \omega \over \pi}
\label{vorden}
\end{equation}
according to Eqs.~(\ref{circ}) and (\ref{flcirc}).  
The transverse size of a vortex is determined by
balancing the competing effects of $- \vec{\nabla} q$ 
and $ - {1 \over m} \vec{\nabla} V$ in Eq.~(\ref{Euler}).
$-\vec{\nabla} q$ tends to increase the transverse size of 
the vortex whereas $ - {1 \over m} \vec{\nabla} V$ tends to 
decrease it assuming, as is the case in superfluid $^4$He, 
that the interparticle interactions are repulsive at short
distances.  The outcome determines the transverse size of 
a vortex to be a characteristic length $\xi$, called the 
`healing length' \cite{Pethik}.  For the interparticle 
potential 
\begin{equation}
V_p(\vec{x} - \vec{x}^\prime) = 
U_0 \delta^3(\vec{x} - \vec{x}^\prime)
\label{deltapot}
\end{equation}
the healing length is 
\begin{equation}
\xi = {1 \over \sqrt{2 m n_0 U_0}}~~~\ .
\label{heal}
\end{equation}
The transverse size of a $l$-vortex, i.e. a vortex that carries 
$l$ units of angular momentum, is $l \xi$.  Indeed the behaviour 
at short distances to the vortex center is the same as in 
Eq. (\ref{example}) with $k$ replaced by $\xi^{-1}$.  The 
cross-sectional area of a $l$-vortex is therefore of order
$\pi \xi^2 l^2$.  Also its energy per unit length \cite{Pethik} 
is proportional to $l^2$.  The vortices repel each other because 
a $l$-vortex has more energy per unit length than $l$ 1-vortices.
The lowest energy configuration for given angular momentum per 
unit area is a triangular lattice of parallel 1-vortices \cite{Tkach}.  
Such triangular vortex arrays were observed in superfluid $^4$He 
\cite{Pack} and in Bose-Einstein condensed gases \cite{Kett}.

\subsubsection{Quasi-collisionless particles}

We finally arrive at the object of our interest, a Bose-Einstein 
condensate of quasi-collisionless particles.  The particles cannot 
be exacly collisionless since they must thermalize to form a BEC 
and they can only thermalize if they interact.  However, the 
interaction by which the particles thermalize can be arbitrarily 
weak since the thermalization may, in principle, occur on an 
arbitrarily long time scale.  In that limit we may set $V = 0$ 
in Eq.~(\ref{Schro}).  For the sake of definiteness we set the 
gravitational field $\Phi = 0$ as well.  We have then
\begin{eqnarray} 
\eta &=& \int d^3r~\Psi^*(- {1 \over 2 m} \nabla^2 
- {\omega \over i} {\partial \over \partial \phi}) \Psi\nonumber\\
&=& {1 \over N} \int d^3r~ n~ [{m \over 2} \vec{v}\cdot\vec{v}
+ m q - m \omega \hat{z}\cdot(\vec{r}\times\vec{v})]~~~\ .
\label{colleta}
\end{eqnarray}
If one approximates the Bose-Einstein condensate as a fluid of 
classical particles (setting $q=0$ and taking $n(\vec{r})$ and 
$\vec{v}(\vec{r})$ to be independent variables), the state of lowest 
$\eta$ is one of rigid rotation with angular velocity $\omega \hat{z}$
and all particles placed as far from the $z$-axis as possible. For 
the reasons stated earlier, this is a good approximation only on 
length scales large compared to the BEC de Broglie wavelength.  To 
obtain the exact BEC state, one must solve the eigenvalue problem
\begin{equation}
(- {1 \over 2 m} \nabla^2 
- \omega {1 \over i} {\partial \over \partial \phi}) \Psi_i 
= \eta_i \Psi_i~~~\ .
\label{eigen}
\end{equation}
The BEC state is then $\Psi_0$ such that $\eta_0 = \min_i \eta_i$.
Since by assumption the system conserves angular momentum, 
the operators $- {1 \over 2 m} \nabla^2$ and 
${1 \over i} {\partial \over \partial \phi}$ are simultaneously 
diagonalizable.  Thus $(i =kl)$
\begin{equation}
\Psi_{kl} (z, \rho, \phi) = A_{kl}(z, \rho) e^{i l \phi}~~,~~
\eta_{kl} = \epsilon_{kl} - \omega l
\label{sep}
\end{equation} 
and 
\begin{equation}
- {1 \over 2 m}({\partial^2 \over \partial z^2} 
+ {1 \over \rho}{\partial \over \partial \rho} \rho 
{\partial \over \partial \rho} ~-~ {l^2 \over \rho^2}) A_{kl}
= \epsilon_{kl} A_{kl}~~\ .
\label{zrho}
\end{equation}
Let us consider the particular example of a BEC contained in a 
cylinder of radius $R$ and height $h$.  In this case, the operators
$- {1 \over 2 m} \nabla^2$, 
${1 \over i} {\partial \over \partial \phi}$ and 
$\left({1 \over i} {\partial \over \partial z}\right)^2$ are 
simultaneously diagonalized by 
\begin{equation}
\Psi_{lpn} = e^{i l \phi} \sin({\pi p \over h} z)
J_l(x_{ln} {\rho \over R})
\label{cylin}
\end{equation}
where $l = 0, \pm 1, \pm 2, ...~~$, $p = 1, 2, 3, ...~~$, 
$n = 1, 2, 3, ...~~$, and $x_{ln}$ is the $n^{\rm th}$ root of 
$J_l(x)$ with $x_{l1} < x_{l2} < x_{l3}< ...~$. Since 
\begin{equation}
\epsilon_{lpn} = {1 \over 2 m}[({\pi p \over h})^2 +
({x_{ln} \over R})^2]
\label{epslpn}
\end{equation}
$\eta_{lpn}$ is minimized by setting $p = n = 1$ and $l = l_0$
where $l_0$ minimizes ${1 \over 2 m}({x_{l1} \over R})^2 - \omega l$.
For large $l$, the first zero of $J_l$ \cite{Abram}
\begin{equation}
x_{l1} \simeq l + 1.85575~l^{1 \over 3} + {\cal O}(l^{-{1 \over 3}})~~\ .
\label{zero1}
\end{equation}
Hence
\begin{equation}
l_0 = m R^2 \omega [1 - 2.47433~(m R^2 \omega)^{-{2 \over 3}} 
+ {\cal O}(m R^2 \omega)^{-{4 \over 3}}]~~~\ .
\label{ello}
\end{equation}
When the BEC is approximated as a fluid of classical particles, 
the BEC state is rigid rotation with all the particles located 
at $\rho = R$, not necessarily in a uniform way.  In the actual 
BEC state the particles are, for large $l$, uniformly located 
just inside the $\rho = R$ surface, in a film of thickness 
$\delta \rho \sim R ~l^{-{2 \over 3}}$.

Unlike the case of superfluid $^4$He, vortices in a collisionless 
BEC attract each other.  Indeed the lowest energy state for given 
total angular momentum $l$ is a single $l$-vortex with transverse 
size as large as possible.   We may imagine turning off the 
interparticle repulsion in superfluid $^4$He placed in a cylindrical 
container.  Starting with a triangular array of $l$ parallel 1-vortices 
but progressively decreasing $U_0$, the vortices grow in transverse 
size till they join into a single $l$-vortex and all matter is
uniformly concentrated near the $\rho = R$ surface.

\subsection{Thermalization and vortex formation}

We emphasized that vortices cannot appear spontaneously in a 
fluid that is described by a (single) wavefunction $\Psi$.  
The Gross-Pitaevskii equation can only describe the motion 
of vortices, not their appearance.  How then do the vortices 
appear?  The vortices appear when the bosons move between 
different particle states, some of which have vortices and 
some of which don't.  When angular momentum is given to a 
BEC that is free of vortices, it will at first remain free 
of vortices even though it carries angular momentum.  The 
vortices only appear when the BEC rethermalizes and the 
particles go to the new lowest energy state consistent 
with the angular momentum the BEC received.  

Consider, for example, a BEC of spin zero particles in 
a cylindrical volume.  The wavefunctions of the particle 
states are given by Eq.~(\ref{cylin}).  The Hamiltonian 
is the sum of free and interacting parts: $H = H_0 + H_1$.
The free Hamiltonian is:
\begin{equation}
H_0 = \sum_{lpn} ~\epsilon_{lpn} a_{lpn}^\dagger a_{lpn}
\label{freeH}
\end{equation} 
where $a_{lpn}$ and $a_{lpn}^\dagger$ are annihilation and 
creation operators satisfying canonical commutation relations
and generating a Fock space in the usual fashion.  We assume 
that the interaction has the general form  
\begin{equation}
H_1 = \sum_{i,i^\prime,i^{\prime\prime},i^{\prime\prime\prime}}~
{1 \over 4} \Lambda_{i^{\prime\prime}~i^{\prime\prime\prime}}^{i~i^\prime}
~a_{i^{\prime\prime\prime}}^\dagger~a_{i^{\prime\prime}}^\dagger~
a_{i^\prime}~a_i
\label{expH1}
\end{equation}
where $i \equiv {lpn}$, $i^\prime \equiv l^\prime p^\prime n^\prime$ 
and so forth, so that the total number of particles
\begin{equation}
N = \sum_{lpn} ~a_{lpn}^\dagger~a_{lpn}
\label{totN}
\end{equation}
is conserved.  In addition we require that
\begin{equation}
\Lambda_{i^{\prime\prime}~i^{\prime\prime\prime}}^{i~i^\prime} = 0
~~~~{\rm unless}~~~ l + l^\prime = l^{\prime\prime} + l^{\prime\prime\prime}
\label{Lcond}
\end{equation}
so that the total angular momentum
\begin{equation}
L = \sum_{lpn}~l~a_{lpn}^\dagger~a_{lpn}
\label{totangmom}
\end{equation}
is conserved as well.  The interaction $H_1$ causes the system to 
thermalize on some time scale $\tau = {1 \over \Gamma}$.  Ref. \cite{therm} 
estimates the thermalization rate $\Gamma$ of cold dark matter axions through 
their $\lambda \phi^4$ and gravitational self-interactions.  The relevant thing 
for our discussion here is only that there is a finite time scale  $\tau = 
{1 \over \Gamma}$ over which the system thermalizes.

Let us suppose that $N$ particles are in thermal equilibrium in the 
cylinder with $\omega = 0$ and temperature $T$ well below the critical 
temperature for Bose-Einstein condensation.  A macroscopically large 
number $N_0$ of particles are in the ground state $(l,p,n)_0 = (0,1,1)$, 
which we label $i=0$ for short. The remaining $N - N_0$ particles are in 
excited ($i \neq 0$) states.  The vorticity of each state equals its $l$ 
quantum number.  The $N_0$ particles in the ground state form a fluid 
with zero vorticity.  Many excited states carry vorticity but their 
occupation numbers are small compared to $N_0$.  The particles in 
excited states merely constitute a gas at temperature $T$.  Let us 
suppose that the fluid is then given some angular momentum.  This can 
be done, for example, by having a large mass $M$ which gravitationally 
attracts the particles in the cylinder go by, producing a time-dependent 
potential energy $V_{\rm ext}(\vec{r},t)$.  We assume for the sake of 
definiteness that the mass $M$ passes by the cylinder on a time scale 
$\tau_M$ which is much shorter than the thermal relaxation time scale 
$\tau$.  While the mass $M$ passes by, each $\phi$ particle stays in 
whatever state it was in to start with since the interaction $H_1$ that 
allows particles to jump between states is, by assumption,  too feeble 
to have any effect on the $\tau_M$ time scale.  The wavefunction of each 
state satisfies the time-dependent Schr\"odinger equation:
\begin{equation}
i \partial_t \Psi_i(\vec{r},t) = [- {1 \over 2m} \nabla^2 
+ V_{\rm ext}(\vec{r},t)] \Psi_i(\vec{r},t)
\label{tdepS}
\end{equation}
with the initial condition 
$\Psi_i(\vec{r}, t = - \infty) = \Psi_i(\vec{r})$.  Although each 
$\Psi_i(\vec{r},t)$ changes in time, for the reasons given earlier, 
its vorticity does not.  Therefore, just after the mass $M$ has passed, 
the macroscopic fluid described by $\Psi_0(\vec{r},t)$ has no vorticity 
although it generally has angular momentum.  After a time of order $\tau$, 
the $N$ particles acquire a thermal distribution, Eq.~(\ref{moddis}) with 
$\sigma = +1$, consistent with the total number of particles $N$, the 
angular momentum $L$ acquired from the passing mass and total energy $E$ 
including some energy acquired from the passing mass.  Assuming the 
temperature is still below the critical temperature for Bose-Einstein 
condensation, a macroscopically large number of particles are in the 
state $(l,p,n)_0^\prime = (l_0^\prime,1,1)$ with $l_0^\prime$ given by
Eq.~(\ref{ello}).  That state describes a fluid which carries a single 
vortex with $l_0^\prime$ units of angular momentum.

\section{Axions, baryons and WIMPs}

In this section we apply the considerations of Section II to dark matter 
axions when they are about to fall into a galactic gravitational potential
well.   We also discuss the behavior, in the presence of dark matter axions, 
of baryons and of a possible ordinary cold dark matter component made of 
weakly interacting massive particles (WIMPs) and/or sterile neutrinos.  
First we discuss the axions by themselves, ignoring the other particles.  

\subsection{Axions}

Axions behave differently from ordinary cold dark matter particles, such 
as WIMPs or sterile neutrinos, on time scales long compared to their 
thermalization time scale $\tau \equiv {1 \over \Gamma}$ because on 
time scales long compared to $\tau$ the axions form a BEC and almost 
all axions go to their lowest energy available state \cite{CABEC,therm}.  
Ordinary cold dark matter particles do not do this.  

It may be useful to clarify the notion of {\it lowest energy available 
state}.  Thermalization involves interactions.  By lowest energy available 
state we mean the lowest energy state that can be reached by the thermalizing 
interactions.  In general the system has states of yet lower energy.  For 
example, and at the risk of stating the obvious, when a beaker of superfluid 
$^4$He is sitting on a table, the condensed atoms are in their lowest energy 
available state.  This is not their absolute lowest energy state since the 
energy of the condensed atoms can be lowered by placing the beaker on the 
floor.  

Axions behave in the same way as ordinary cold dark matter on time scales 
short compared to their thermalization time scale $\tau$ \cite{CABEC}.  So, 
to make a distinction between axions and ordinary cold dark matter it is 
necessary to observe the dark matter on time scales long compared to $\tau$.  
The critical question is then: what is the thermalization time scale $\tau$?

\subsubsection{Axion thermalization}

The relaxation rate of axions through gravitational self-interactions 
is of order \cite{CABEC,therm,Saik}
\begin{equation}
\Gamma \sim 4 \pi G n m^2 \ell^2
\label{relax}
\end{equation}
where $n$ and $m$ are their density and mass, and 
$\ell \equiv {1 \over \delta p}$ their correlation length. 
$\delta p$ is their momentum dispersion.  A heuristic derivation
of Eq.~(\ref{relax}) is as follows.  If the axions have density $n$
and correlation length $\ell$, they produce gravitational fields of 
order $g \sim 4 \pi G n m \ell$.  Those fields completely change the 
typical momentum $\delta p$ of axions in a time ${\delta p \over g m}$.  
$\Gamma$ is the inverse of that time.  To estimate the axion relaxation 
time today, let us substitute $n m \simeq 0.23 \cdot 10^{-29}$ gr/cc
(the average dark matter density today), $m \simeq 10^{-5}$ eV (a typical 
mass for dark matter axions) and 
$\ell \simeq 2 \cdot 10^{-7} {\rm sec} {{\rm GeV} \over 10^{-4} {\rm eV}}
= 0.6 \cdot 10^{17} {\rm cm}$ (the horizon during the QCD phase transition, 
stretched by the universe's expansion until today).  This yields a relaxation 
time $\tau$ of order $10^5$ years, much shorter than the present age of the 
universe.  So dark matter axions formed a BEC a long time ago already.  It 
is found in refs. \cite{CABEC,therm} that the axions first thermalize and 
form a BEC when the photon temperature is approximately 
500 eV$\left({f_a \over 10^{12}~{\rm GeV}}\right)^{1 \over 2}$ where $f_a$
is the axion decay constant. After the axions form a BEC their correlation 
length $\ell$ increases until it is of order the horizon since the BEC size 
is limited only by causality.  

It may seem surprising that axions thermalize as a result of their 
gravitational self-interactions since gravitational interactions among 
particles are usually negligible.  Dark matter axions are an exception 
because the axions occupy in huge numbers a small number of states 
(the typical quantum state occupation number is $10^{61}$) and those 
states have enormous correlation lengths, as was just discussed.

It has been claimed \cite{CABEC, case,therm} that the dark matter is axions, 
at least in part, because axions explain the occurrence of caustic rings of 
dark matter in galactic halos.  For the explanation to succeed it is necessary 
that the axions that are about to fall onto a galaxy thermalize sufficiently 
fast that they almost all go to the lowest energy available state consistent 
with the angular momentum they acquired from neighboring protogalaxies by tidal
torquing \cite{TTT}.  Heuristically, the condition is \cite{therm} 
\begin{equation} 
4 \pi G n m^2 \ell > \dot{p} = m \dot{v}
\label{galcon} 
\end{equation} 
where $\dot{v}$ is the acceleration necessary for the axions to remain in the 
lowest energy state as the tidal torque is applied.  Here $\ell$ must be taken 
to be of order the size of the system, i.e. some fraction of the distance between
neighboring protogalaxies.  It was found in ref. \cite{therm} that the inequality 
(\ref{galcon}) is satisfied by a factor of order 30 - i.e.  that its LHS is of 
order 30 times larger than its RHS - independently of the system size.

\subsubsection{Caustic rings}

The evidence for caustic rings is summarized in ref. \cite{MWhalo}.  It 
is accounted for if the angular momentum distribution of the dark matter 
particles on the turnaround sphere of a galaxy is given by 
\begin{equation}
\vec{l}(\hat{n}, t) = m~j_{\rm max}~\hat{n} \times (\hat{z} \times \hat{n})~
{R(t)^2 \over t}
\label{spec}
\end{equation}
where $t$ is time since the Big Bang, $R(t)$ is the radius of the turnaround 
sphere, $\hat{z}$ the galactic rotation axis, $\hat{n}$ the unit vector pointing 
to an arbitrary point on the turnaround sphere, and $j_{\rm max}$ a dimensionless
parameter that characterizes the amount of angular momentum the particular
galaxy has.  The turnaround sphere is defined as the locus of particles which 
have zero radial velocity with respect to the galactic center for the first time, 
their outward Hubble flow having just been arrested by the gravitational pull of 
the galaxy.  Eq.~(\ref{spec}) states that the particles on the turnaround sphere 
rotate rigidly with angular velocity vector $\vec{\omega} = {j_{\rm max} \over t}
\hat{z}$.  The time-dependence of the turnaround radius is predicted by the
self-similar infall model \cite{FGB} to be 
$R(t) \propto t^{{2 \over 3} + {2 \over 9 \epsilon}}$.  The parameter $\epsilon$ 
is related to the slope of the evolved power spectrum of density perturbations 
on galaxy scales \cite{Dor}.  This implies that $\epsilon$ is in the range
0.25 to 0.35 \cite{STW}.  The evidence for caustic rings is consistent with 
that particular range of values of $\epsilon$.  

Each property of the angular momentum distribution given in Eq.~(\ref{spec}) 
maps onto an observable property of the inner caustics of galactic halos: the 
rigid rotation implied by the factor $\hat{n} \times (\hat{z} \times \hat{n})$ 
causes the inner caustics to be rings of the type described in refs. 
\cite{crdm,sing,inner}, the value of $j_{\rm max}$ determines their overall 
size, and the ${R(t)^2 \over t}$ time dependence causes, in the stated $\epsilon$
range, the caustic radii $a_n$ to be proportional to $1/n$ ($n = 1, 2, 3 ...$).  
The prediction for the caustic radii is
\begin{equation}
a_n \simeq {40~{\rm kpc} \over n}~ 
\left({v_{\rm rot} \over 220~{\rm km/s}}\right)~
\left({j_{\rm max} \over 0.18}\right)
\label{crr}
\end{equation}
where $v_{\rm rot}$ is the galactic rotation velocity.  To account for 
the evidence for caustic rings, axions must explain Eq.~(\ref{spec}) in 
all its aspects.  We now show, elaborating the arguments originally given 
in ref.\cite{case}, that axions do in fact account for each factor on the 
RHS of Eq.~(\ref{spec}).

\subsubsection{$\hat{n}\times(\hat{z}\times\hat{n})$}

Consider a comoving spherical volume of radius $S(t)$ centered on a
protogalaxy.  At early times $S(t) = a(t) S$ where $a(t)$ is the 
cosmological scale factor.  At later times $S(t)$ deviates from 
Hubble flow as a result of the gravitational pull of the protogalactic
overdensity.  At some point it reaches its maximum value.  At that moment
it equals the galactic turnaround radius.  $S$ is taken to be of order 
but smaller than the distance to the nearest protogalaxy of comparable 
size, say one third of that distance.  In the absence of angular 
momentum, the axions have a purely radial motion described by a 
wavefuntion $\Psi(r,t) = U(r,t)$ where $r$ is the radial coordinate 
relative to the center of the sphere.  When angular momentum is included 
the radial motion is modified at small radii by the introduction of an 
angular momentum barrier. This modification of the radial motion is 
relatively unimportant and we neglect it.  The wavefunctions of the 
states that the axions occupy are thus taken to be
\begin{equation}
\Psi_{l,p}(r,\theta,\phi,t) = U(r,t) A_{l,p}(\theta,\phi)
\label{posswav}
\end{equation}
where $\theta$ and $\phi$ are the usual spherical angular 
coordinates ($0 \leq \theta \leq \pi$, $0 \leq \phi < 2 \pi$), 
and $l$ and $p$ are quantum numbers.  $l$ is as before the 
eigenvalue of the $z$-component of angular momentum.  The 
$z$-direction is the direction of the total angular momentum 
acquired inside the sphere as a result of tidal torquing.  
We will see below that that direction is time independent.  
$p$ is an additional quantum number, associated with motion in 
$\theta$.  We normalize $U(r,t)$ and the various $A(\theta,\phi)$ 
such that 
\begin{equation}
\int_0^{S(t)}~r^2~dr~|U(r,t)|^2 = 
\int_0^\pi ~\sin\theta~d\theta~\int_0^{2 \pi}~d\phi~|A(\theta,\phi)|^2 
= 1~~\ .
\label{norm}
\end{equation}
We suppress the quantum numbers $l$ and $p$ henceforth.

According to Eq.~(\ref{galcon}) the axions thermalize on a time 
scale $\tau$ that is short compared to the age of the universe.
Hence we expect most axions to keep moving to the state of lowest 
$~\eta = \epsilon - \omega(t) l~~$.  The angular frequency $\omega$ 
is time dependent since the angular momentum is growing by tidal 
torquing and the moment of inertia is increasing due to the expansion 
of the volume under consideration.  We have
\begin{eqnarray}
\epsilon &=& \int_{r < S(t)}~d^3x~ \Psi^* [- {1 \over 2 m} \nabla^2 
+ m \Phi(r,t)] \Psi \nonumber\\
&=& \int_0^{S(t)}~r^2 dr~U^*[- {1 \over 2 m r^2} 
{\partial \over \partial r} r^2 {\partial \over \partial r}
~+~ m \Phi(r,t)] U \nonumber\\
&+& {1 \over 2 I(t)} \int~d\Omega~A^*
[- {1 \over \sin\theta} 
{\partial \over \partial \theta} \sin\theta 
{\partial \over \partial \theta}
~-~{1 \over \sin^2\theta} 
{\partial^2 \over \partial \phi^2} ] A
\label{axeps} 
\end{eqnarray}
where 
\begin{equation}
{1 \over I(t)} = {1 \over m} \int_0^{S(t)}~dr~|U(r,t)|^2~~~\ .
\label{inmom}
\end{equation}
$I(t)$ is similar to a moment of inertia but differs from the usual 
definition because the volume to which it refers is not rotating like 
a rigid body in three dimensions.  Eq.~(\ref{posswav}) implies instead
that each spherical shell of that volume rotates with the same angular 
momentum distribution.  The associated angular velocities vary with shell
radius $r$ as $r^{-2}$. The inner shells rotate faster than the outer 
shells because all shells have the same angular momentum distribution.
  
We take the gravitational potential $\Phi$ to be spherically 
symmetric.  The first term on the RHS of Eq.~(\ref{axeps}) is 
then independent of the angular variables and irrelevant to what 
follows.  We will ignore it henceforth.  Since
\begin{equation}
l = \int~d\Omega~A^*~{1 \over i} {\partial \over \partial \phi} A
\label{axell}
\end{equation}
we have 
\begin{eqnarray}
\eta &=& \int d\Omega~A^* [{1 \over 2 I(t)}( - {1 \over \sin\theta}
{\partial \over \partial \theta} \sin\theta {\partial \over \partial \theta}
- {1 \over \sin^2\theta} {\partial^2 \over \partial \phi^2}) 
- \omega(t) {1 \over i} {\partial \over \partial \phi}] A \nonumber\\
&=& \int d\Omega~A^* [{1 \over 2 I(t)} \left(-{1 \over \sin\theta}
{\partial \over \partial \theta} \sin\theta {\partial \over \partial \theta}
+ {1 \over \sin^2\theta} ({1 \over i} {\partial \over \partial \phi}
- \omega(t) I(t) \sin^2\theta)^2\right) \nonumber\\
&~&~~~~- {1 \over 2} \omega^2(t) I(t) \sin^2\theta] A\nonumber\\
&=& \int d\Omega~ [{1 \over 2 I(t)} |{d A \over d \theta}|^2 
- {1 \over 2} \omega^2(t) I(t) \sin^2\theta |A|^2\nonumber\\
&~&~~~~
+ {1 \over 2 I(t) \sin^2\theta}|({1 \over i}{\partial \over \partial \phi}
- \omega(t) I(t) \sin^2\theta) A|^2]~~~\ .
\label{axeta}
\end{eqnarray}
The $\phi$ dependence of $A$ that minimizes $\eta$ is  
\begin{equation}
A(\theta,\phi,t) = \Theta(\theta,t) e^{i \omega(t) I(t) \sin^2\theta~\phi}~~~\ .
\label{Amin}
\end{equation}
However that exact $\phi$ dependence is not allowed because the wavefunction 
must be single-valued.  Instead we have 
\begin{equation}
A(\theta,\phi,t) \simeq \Theta(\theta,t) e^{i \omega(t) I(t) \sin^2\theta~\phi} 
\label{Aminp}
\end{equation}
by which we mean that $A(\theta,\phi,t)$ is as given in Eq.~(\ref{Amin}) 
except for the insertion of small defects (vortices) that allow $A$ to 
be single-valued.  The vortices are discussed below.  

After Eq.~(\ref{Aminp}) is satisfied, we have 
\begin{equation}
\eta \simeq \int d\Omega~ 
[{1 \over 2 I(t)} |{d \Theta \over d \theta}|^2
- {1 \over 2} \omega^2(t) I(t) \sin^2\theta |\Theta|^2]~~~\ .
\label{etaTh}
\end{equation}
$\eta$ is further minimized by having $\Theta$ peaked at 
$\theta = {\pi \over 2}$, i.e. at the equator.  The width of the 
peak is of order $\delta \theta \sim {1 \over \sqrt{\omega I}}$.  
Eq.~(\ref{Aminp}) shows that $L \equiv \omega I$ is the angular 
momentum per particle in the galactic plane.  A typical value is 
\begin{equation}
L \sim (500~{{\rm km} \over {\rm s}})(10~{\rm kpc})~m 
\simeq~2.6 \cdot 10^{19} \left({m \over 10^{-5}~{\rm eV}}\right)~~\ .
\label{estim}
\end{equation}
Therefore in their state of lowest $\eta$ the axions are almost all 
within a very small angular distance, of order $10^{-10}$ radians, 
from the galactic plane.

The state just described is the state most axions would be in after a 
sufficiently long period of thermalization.  Because the thermalization 
criterion of Eq.~(\ref{galcon}) is only satisfied by a factor of order 
30 we expect that, although the axions start to move towards the equator, 
there is not enough time for all the axions to get localized there.  We 
expect the system to behave as follows.  As the axions acquire angular 
momentum they go to a state, described by Eqs.~(\ref{posswav}) and 
(\ref{Aminp}), in which each spherical shell rotates rigidly with  
angular velocity proportional to $r^{-2}$ where $r$ is the shell 
radius.  The axion velocity field is  
\begin{equation} 
\vec{v} \simeq v_r \hat{r} + v_\theta \hat{\theta} + 
{1 \over m r} L(t) \sin\theta \hat{\phi}~~\ . 
\label{axvel}
\end{equation} 
The $~\simeq~$ sign indicates that the LHS and RHS equal each other except 
for the presence of vortices.  The vortices have direction and density per
unit surface given by [see Eqs.~(\ref{circ}) and (\ref{flcirc})] 
\begin{equation} 
{m \over 2 \pi} \vec{\nabla}\times \vec{v} \simeq
\vec{\nabla}\times \left({L(t) \over m r} \sin\theta \hat{\phi}\right)
= {L(t) \over \pi r^2} \cos\theta~\hat{r}~~~\ . 
\label{vortden} 
\end{equation} 
They point in the radial direction, and are more dense near the poles 
than near the equator.  The total vortex number penetrating the northern 
hemisphere is
\begin{equation} 
\int_0^{2 \pi} d\phi~\int_0^{\pi \over 2}~\sin\theta~d\theta~r^2~ 
{m \over 2 \pi}\vec{\nabla}\times\vec{v} = L~~~\ .
\label{totvort} 
\end{equation} 
As discussed in Section II, axion vortices attract each other.  When two 
vortices combine, their diameters are added.  (Two vortices of equal diameter, 
and hence of equal cross-sectional area, combine into a vortex with four times 
that cross-sectional area).  Assuming that a fraction of order one of all the
vortices combine with one another, a huge vortex appears along the $\hat{z}$
axis.  We will refer to it as the `big vortex'.  The intersection of the big
vortex with the galactic plane is a circle whose radius 
$a^\prime \simeq {L^\prime \over k}$ where $k$ is the momentum of axions in 
the equatorial plane at their closest approach to the galactic center and 
$L^\prime$ is the angular momentum carried by the big vortex.  The distance 
of closest approach to the galactic center of axions in the equatorial plane 
is $a= {L \over k}$.  $a$ is also the radius of the caustic ring made by the 
axions as they fall through the galaxy for the first time.   Because of 
incomplete thermalization, we expect that some fraction of the vortices
have not joined the big vortex, implying that $L^\prime < L$ and hence 
$a^\prime < a$.

The factor $\hat{n}\times(\hat{z}\times\hat{n})$ in the angular momentum
distribution on the turnaround sphere, Eq.~(\ref{spec}), is thus accounted for by
the fact that the axions on the turnaround sphere rotate rigidly.  After turnaround
there is not enough time for further thermalization and whatever further
thermalization may occur would not make an appreciable difference.  Thus, after
turnaround, the axions fall in and out of the galaxy like ordinary cold
collisionless particles but they do so with net overall rotation whereas ordinary
cold dark matter falls in with an irrotational velocity field \cite{inner}.

Why and how the angular momentum distribution of Eq.~(\ref{spec}) yields 
caustic rings is explained in refs.~\cite{sing,MWhalo}.  However those 
papers, written before the discovery of Bose-Einstein condensation of 
dark matter axions \cite{CABEC}, assume that the infall is isotropic, 
i.e. that the mass falling onto the galaxy per unit time and unit solid 
angle ${dM \over d\Omega dt}$ does not depend on $\theta$ (nor on $\phi$).  
The above discussion suggests that this assumption should be modified 
since a big vortex is now expected along the $\hat{z}$ axis, implying 
that the infall rate is suppressed near $\theta = 0$ and $\pi$.  In 
the self-similar infall model the total mass $M$ of the halo grows as 
$t^{2 \over 3 \epsilon}$ \cite{FGB}.  Therefore 
${d M \over d \Omega dt} = {M \over 6 \pi \epsilon t}$ in the isotropic 
case.  We replace this with
\begin{equation}
{dM \over d \Omega dt} (\theta, t) 
= N_\upsilon (\sin\theta)^\upsilon {M \over 6 \pi \epsilon t}
\label{infr}
\end{equation}
where $\upsilon$ (lower case upsilon, not to be confused with $v$
the magnitude of velocity) is a parameter describing the size of 
the big vortex. The normalization factor 
\begin{equation}
N_\upsilon = {\Gamma(\upsilon+2) \over 2^\upsilon 
(\Gamma({\upsilon \over 2} + 1))^2}
\label{Np}
\end{equation}
is such that the total infall rate, integrated over solid angle, 
remains the same as before.  The new model for the infall rate, 
Eq.~(\ref{infr}), does not change the prediction that the inner 
caustics are rings nor the prediction, Eq.~(\ref{crr}), for the 
caustic ring radii \cite{NBPS}.  It does however imply (for large 
$\upsilon$) that the caustic rings are more prominent than in the 
isotropic infall case since the axions fall in preferentially along 
the galactic plane.  This will be discussed in Section IV.

\subsubsection{${R(t)^2 \over t}$}

We now show, repeating the argument of ref. \cite{case}, that the
${R(t)^2 \over t}$ time dependence on the RHS of Eq.~(\ref{spec}) 
follows from tidal torque theory in linear order of perturbation 
theory.  Consider again the comoving sphere of radius $S(t)$ 
introduced above.  The total gravitational torque applied to the 
volume $V(t)$ of the sphere is
\begin{equation} 
\vec{\tau}(t) = \int_{V(t)} d^3r
~\delta\rho(\vec{r}, t)~\vec{r}\times (-\vec{\nabla} \phi(\vec{r}, t))
\label{torq}
\end{equation}
where $\delta\rho(\vec{r}, t) = \rho(\vec{r}, t) - \rho_0(t)$ is
the density perturbation.  $\rho_0(t)$ is the unperturbed density.
In first order of perturbation theory, the gravitational potential 
does not depend on time when expressed in terms of comoving coordinates, 
i.e. $\phi(\vec{r} = a(t) \vec{x}, t) = \phi(\vec{x})$.  Moreover
$\delta(\vec{r}, t) \equiv {\delta \rho(\vec{r}, t) \over \rho_0(t)}$
has the form $\delta(\vec{r} = a(t) \vec{x}, t) = a(t) \delta(\vec{x})$.
Hence
\begin{equation} 
\vec{\tau}(t) = \rho_0(t) a(t)^4 \int_V d^3x
~\delta(\vec{x})~\vec{x} \times (- \vec{\nabla}_x \phi(\vec{x}))~~~\ .
\label{tt}
\end{equation}
Eq.~(\ref{tt}) shows that the direction of the torque is time 
independent.  Hence the rotation axis is time independent, as in 
Eq.~(\ref{spec}).  Furthermore, since $\rho_0(t) \propto a(t)^{-3}$, 
$\tau(t) \propto a(t) \propto t^{2 \over 3}$ and hence the angular 
momentum increases with time proportionally to $t^{5 \over 3}$.  
Since $R(t) \propto t^{{2 \over 3} + {2 \over 9 \epsilon}}$, tidal 
torque theory predicts the time dependence of Eq.~(\ref{spec}) 
provided $\epsilon = 0.33$. This value of $\epsilon$ is in the 
range, $0.25 < \epsilon < 0.35$, predicted by the evolved spectrum 
of density perturbations and supported by the evidence for caustic 
rings.  So the time dependence in Eq.~(\ref{spec}) is accounted for.

\subsubsection{$j_{\rm max}$}

Here we compare the average value of $j_{\rm max}$ implied by the 
evidence for caustic rings with the amount of angular momentum 
expected from tidal torquing. The amount of angular momentum 
acquired by a galaxy through tidal torquing can be reliably 
estimated by numerical simulation because it does not depend 
on any small feature of the initial mass configuration, so that 
the resolution of present simulations is not an issue in this case.  
The amount of galactic angular momentum is usually given in terms of 
the dimensionless quantity \cite{Peeb}
\begin{equation}
\lambda \equiv {{\cal L} |{\cal E}|^{1 \over 2} 
\over G {\cal M}^{5 \over 2}}~~\ ,
\label{lambda}
\end{equation}
where ${\cal L}$ is the angular momentum of the galaxy, ${\cal M}$ 
its mass and ${\cal E}$ its net mechanical (kinetic plus gravitational 
potential) energy.  $\lambda$ was found in numerical simulations to 
have a broad distribution with median value 0.05 \cite{Efst}.  $\lambda$ 
may also be estimated from observations of the luminous matter by making 
some assumptions, in particular the assumption that the angular momentum 
per unit mass of the disk and the halo are equal.  Using such methods, 
Hernandez et al. \cite{Hern} derived the $\lambda$ distribution of a 
large sample of spiral galaxies from the Sloan Digital Sky Survey and 
found it to be consistent with the expectations from numerical simulations.

On the other hand, in the caustic ring model, the dimensionless 
measure of galactic angular momentum is $j_{\rm max}$.  The evidence 
for caustic rings implies that the $j_{\rm max}$ distribution is peaked 
at $j_{\rm max} \simeq$ 0.18. In case of isotropic infall ($\upsilon =0$), 
the relationship between $j_{\rm max}$ and $\lambda$ is \cite{case}
\begin{equation}
\lambda = \sqrt{6 \over 5 - 3 \epsilon}~
{8 \over 10 + 3 \epsilon}~
{1 \over \pi}~j_{\rm max}~~~\ .
\label{rel}
\end{equation}
For $\epsilon$ = 0.33, Eq.~(\ref{rel}) implies $\lambda/j_{\rm max}$ = 0.283. 
Hence there is excellent agreement between $j_{\rm max} \simeq 0.18$ and 
$\lambda \sim 0.05$ when $\upsilon=0$.  That the agreement is so good is 
likely somewhat fortuitous since neither the $\lambda$ nor the $j_{\rm max}$ 
distribution is very well established.  Also, because the extent of galactic 
halos is not uniformly agreed upon, there is some amibiguity in the definition 
of the quantities, ${\cal L}$, ${\cal E}$ and ${\cal M}$, that enter $\lambda$.  
In deriving Eq.~(\ref{rel}), the halo was taken to extend all the way to the 
turnaround radius \cite{case}.  In obtaining the $\lambda$ distribution from 
numerical simulations, a definition of a galactic halo convenient in numerical 
simulations is used \cite{Efst}.  The two definitions are not clearly equivalent.  
All together the agreement between $\lambda$ and $j_{\rm max}$ is significant 
only within some factor of order one, perhaps as large as 2.  

For ${\cal \upsilon} \neq 0$, the relationship between $\lambda$ and $j_{\rm max}$ 
is more difficult to derive because of the lack of spherical symmetry.   A 
calculation implies that the RHS of Eq.~(\ref{rel}) is multiplied by a factor 
of order ${1 + \upsilon/2 \over 1 + \upsilon/3}$.  In a future publication, 
we will justify this factor and produce refinements to it \cite{NBPS}.  If 
$\upsilon$ is much larger than one, the RHS of Eq.~(\ref{rel}) is multiplied 
by 3/2 so that ${\lambda \over j_{\rm max}} \simeq 0.426$ for $\epsilon$ =
0.33.  The values $\lambda \sim 0.05$ and $j_{\rm max} \simeq 0.18$ are then 
consistent at the 50\% level only.

\subsection{Baryons and WIMPs}

The gravitational forces produced by the axion BEC act not only 
on the axions themselves but also on all other particles present.
In particular, the axion BEC interacts gravitationally with baryons, 
and with WIMPs if WIMPs are present.  The condition for baryons/WIMPs 
to acquire net overall rotation by gravitational interaction with the 
axion BEC is heuristically 
\begin{equation}
4 \pi G n m m^\prime \ell > m^\prime~\dot{v}~~\ ,
\label{bcon}
\end{equation}
where $m^\prime$ is the baryon/WIMP mass.  The accelerations
$\dot{v}$ necessary to acquire net overall rotation are the same
for baryons/WIMPs as for axions.  Since $m^\prime$ cancels out of the 
inequality (\ref{bcon}), conditions (\ref{bcon}) and (\ref{galcon}) are 
the same.  It was found in ref. \cite{therm} that, if the dark matter 
is entirely axions, the inequality (\ref{galcon}) is satisfied by a 
factor of order 30.  Therefore conditions (\ref{galcon})and (\ref{bcon}) 
are equivalent to 
\begin{equation}
n~m \gtrsim {1 \over 30}~\rho_{\rm DM}
\label{axcon}
\end{equation}
where $\rho_{\rm DM}$ is the total cold dark matter density.  If the 
axion fraction of cold dark matter is larger than of order 3\%, we 
expect the axion BEC to acquire net overall rotation and to entrain 
the baryons and WIMPs along.

The baryons and WIMPs do not form a BEC but, by being in thermal 
contact with the axion BEC, they behave in very much the same way.  
Indeed, thermal contact between baryons/WIMPs and axions implies 
that they have the same temperature $T$ and the same angular 
velocity $\omega$, as was discussed in subsection II.A.  The 
temperature of axions is certainly smaller than the typical 
kinetic energy of axions in a galactic halo because axions 
with larger kinetic energy would escape.  Since the 
typical halo velocity is $v \sim 10^{-3} c$,
\begin{equation}
T \lesssim  {1 \over 2} m v^2 \simeq 
6 \cdot 10^{-8}~{\rm K} \left({m \over 10^{-5}~{\rm eV}}\right)~~~\ .
\label{cold}
\end{equation}
The velocity dispersion of the baryons and WIMPs, being much heavier 
than the axions but at the same temperature, is tiny: less than 30 cm/s 
if $m^\prime \geq$ 1 GeV and $m < 10^{-3}$ eV.  The temperature of baryons 
and WIMPs is thus effectively zero.  Baryons and WIMPs are therefore in 
their own state of lowest available $\eta = \epsilon - \omega l$, with 
the same $\omega$ as the axions.  That state may be derived by the same 
methods as we used for the axions in Section II.A.  The outcome is that 
the baryons/WIMPs are in a state of rigid rotation (again not in the 
three dimensional sense, but in the sense that each spherical shell 
rotates rigidly with angular velocity proportional to $r^{-2}$ where 
$r$ is the shell's radius) with the same velocity field, Eq.~(\ref{axvel}), 
as the axion BEC.  The underlying reason for this outcome is simple.  If 
the baryons and WIMPs were not locally at rest with respect to the axion 
BEC, entropy could be generated by bringing them to such a state.  Furthermore, 
as was the case for the axions, the lowest $\eta$ state is one where all the 
baryons and WIMPs are near the equator.  We assume again, as we did for axions, 
that the baryons/WIMPs start to move towards the equator but that there is not 
enough time for them to all get there.  Thus the baryon/WIMP infall rate is 
taken to have the same functional dependence on $t$ and $\theta$ as for axions 
\begin{equation}
{d M^\prime \over d \Omega dt} = N_{\upsilon^\prime} 
(\sin \theta)^{\upsilon^\prime} ~{M^\prime \over 6 \pi \epsilon t}
\label{infrp}
\end{equation}
but we allow a different value $\upsilon^\prime$ for the $\upsilon$
parameter.  Indeed we expect that the axions move to the equator first 
and that the baryons/WIMPs follow them there.  Since the baryons/WIMPs 
are locally at rest with respect to the axion BEC, their angular momentum 
distribution on the turnaround sphere is the same as in Eq.~(\ref{spec}) 
but with $m$ replaced by $m^\prime$:
\begin{equation}
\vec{l}^\prime(\hat{n},t) = m^\prime~j_{\rm max}~ 
\hat{n} \times (\hat{z} \times \hat{n})~{R(t)^2 \over t}~~~\ .
\label{specp}
\end{equation}
Since the WIMPs fall in with the same initial velocity distribution as 
the axions, they move in the same way after falling onto the galaxy and 
produce the same caustic structures.  The baryons also fall in the same 
way initially, but being collisionfull, separate from the axions and WIMPs 
after shell crossing starts.

\section{Comparison with observations}

Here we discuss two observations that appear related to the physics 
described in Section III.  The first is a measurement of the angular 
momentum distribution of baryonic matter in dwarf galaxies by van den Bosch 
et al. \cite{Bosch}.  The second is the typical size of the observed rises 
in the Milky Way rotation curve, compared to the prediction from caustic 
rings of dark matter \cite{MWcr}.

\subsection{Baryonic angular momentum distribution}

In the first part of this subsection we recount a discrepancy between the 
observed and predicted angular momentum distributions of baryons in galaxies 
if the dark matter is ordinary cold dark matter, such as WIMPs.  In the 
second part we show how the discrepancy is resolved if the dark matter is 
axions.

\subsubsection{If the dark matter is all WIMPs}

Since we assume in this subsection that none of the dark matter is axions,
the considerations of Section III do not apply.

By the principle of equivalence, tidal torquing gives the same amount and 
the same distribution of specific angular momentum (i.e. angular momentum 
per unit mass) to baryons and to dark matter before they fall onto galactic 
halos.  Let us assume, to start with, that the individual angular momentum 
of each particle is conserved from its turnaround till today.  In that case 
the observed amount and distribution of baryonic specific angular momentum 
is the same as predicted for dark matter by numerical simulations.  It was 
mentioned already in subsection III.A.5 that the {\it amount} of specific 
angular momentum observed in the baryonic components of disk galaxies is 
consistent with the amount expected from numerical simulations, lending 
support to the hypothesis that the angular momentum of each particle is 
conserved.  However, the observed specific angular momentum {\it distribution} 
of baryons in disk galaxies differs markedly from that predicted by numerical 
simulations for WIMP dark matter.  The predicted distribution has many more 
particles with low specific angular momentum than the observed distribution 
and a compensating (to keep the average the same) population of particles 
with much higher specific angular momentum. The simulations predict too 
high a concentration of baryons at the centers of galaxies.  

At first it may appear that the solution to this discrepancy is simply 
to abandon the notion that the angular momentum of individual particles 
is conserved after they have fallen onto the galaxy.  However, when the 
processes that allow angular momentum exchange are modeled, it is found 
that they aggravate the discrepancy rather than resolve it.  Frictional 
forces among baryons have the general effect of removing angular momentum 
from baryons that have little angular momentum and transferring it to 
baryons that have a lot.  Dynamical friction of dark matter on clumps of 
baryonic matter has the general effect of tranferring angular momentum 
from the baryons to the dark matter.  Both processes tend to concentrate 
baryons at galactic centers even more, aggravating the discrepancy \cite{Nava}.  
The discrepancy is commonly referred to as the `galactic angular momentum 
problem'.  See ref. \cite{Burk} for a review.

The problem is thrown into sharp relief by comparing the universal angular 
momentum distribution obtained by Bullock et al. from numerical simulations 
\cite{Bull} with the observed angular momentum distribution of baryons in 
dwarf galaxies \cite{Bosch}.  Bullock et al. found that the specific angular 
momentum distributions of the galaxies in simulations are all well fitted by 
a single two parameter function:
\begin{equation}
{dM \over dl} = {\mu M_v l_0 \over (l_0 + l)^2}~~~~~~~~{\rm for} 
~~~~0 \leq l \leq l_{\rm max} = {l_0 \over \mu - 1}
\label{numdis}
\end{equation}
where $\mu > 1$, and $M_v$ is the halo's virial mass.  Each galaxy has its 
own value of $\mu$ and $l_{\rm max}$.  The distribution of 
$\log_{10}(\mu - 1)$ values for the galaxies in the simulations is nearly 
Gaussian with average  -0.6 and standard deviation 0.4, implying that 90\% 
of halos have $0.06 < \mu - 1 < 1.0$.  The median $\mu$ value is 1.25.  The 
ratio of the average specific angular momentum $l_{\rm av}$ to the maximum 
specific angular momentum is given in terms of the parameter $\mu$ by 
\begin{equation}
{l_{\rm av} \over l_{\rm max}} = (\mu - 1) 
\left[ \mu \ln\left({\mu \over \mu - 1}\right) - 1 \right]~~~\ .
\label{ratio}
\end{equation}
The broad distribution of $\mu$ values implies a correspondingly 
broad distribution of ${l_{\rm av} \over l_{\rm max}}$ values with 
average near 0.25 \cite{Bosch}.

van den Bosch et al. \cite{Bosch} derived the baryonic angular 
momentum distribution of fourteen dwarf galaxies from observations
by Swaters \cite{Swat}.  The distributions are shown in Fig. I which 
is a reproduction of the relevant figure in ref. \cite{Bosch}.  The 
prediction of Eq.~(\ref{numdis}) with $\mu = 1.25$, the median value, 
is shown as a solid line in each panel. The observed distributions 
are markedly different from the prediction of Eq.~(\ref{numdis}).  
Perhaps the most striking difference is that Eq.~(\ref{numdis}) 
predicts ${d M \over d l}$ to be maximum at $l =0$ whereas the 
observed distributions appear to go to zero at $l = 0$ and have 
their maxima around $l = l_{\rm av}$.  Another striking difference, 
pointed out in ref. \cite{Bosch},  is that the observed values of 
${l_{\rm av} \over l_{\rm max}}$ are strongly peaked near 0.375.  This 
is apparent from the fact that many of the distributions in Fig. I end 
at $l_{\rm max} \simeq 2.6~l_{\rm av}$.  As mentioned, the numerical 
simulations predict ${l_{\rm av} \over l_{\rm max}}$ to have a broad 
distribution with median value 0.25.  If this were so, the distributions 
in Fig. I would end at a wide variety of ${l_{\rm max} \over l_{\rm av}}$ 
values and half of these values would be larger than 4.

\subsubsection{If the dark matter is axions, at least in part}

As described in Section III, the angular momentum distributions of the 
baryons and WIMPs are modified by gravitational interactions with the 
axion BEC.  The outcome is Eq.~(\ref{specp}) for the angular momentum 
distribution on the turnaround sphere and Eq.~(\ref{infrp}) for the 
infall rate.  We assume that the angular momentum of each particle is 
conserved after it crosses the turnaround sphere. Eq.~(\ref{specp}) 
implies for the angular momentum distribution on the turnaround 
sphere at time $t$
\begin{equation}
l(\theta,t) = \hat{z}\cdot\vec{l}(\vec{n},t) = 
l_{\rm max}~(\sin\theta)^2 \left({t \over t_0}\right)^{5 \over 3}
\label{lftt}
\end{equation}
where $t_0$ is the present age of the universe and 
\begin{equation}
l_{\rm max} = m~~j_{\rm max}~{R_0^2 \over t_0}
\label{lmax}
\end{equation}
is the angular momentum of particles falling in along the galactic 
plane today.  $R_0$ is the present turnaround radius.  We are removing 
the primes from $\vec{l}^\prime$, $m^\prime$, $M^\prime$ and so forth 
to avoid cluttering the equations unnecessarily.  To obtain Eq.~(\ref{lftt}) 
from Eq.~(\ref{specp}) we used the fact that $l \propto t^{5 \over 3}$; 
see subsection III.A.4.  The angular momentum distribution today is 
\begin{equation}
{dM \over dl}(l) = \int d\Omega~\int_0^{t_0} dt~
{dM \over d\Omega dt}(\theta,t) ~\delta(l - l(\theta,t))~~~\ .
\label{prin}
\end{equation}
Substituting Eqs.~(\ref{infrp}) and (\ref{lftt}) and carrying 
out the $t$ integration, one finds for $\epsilon =1/3$
\begin{equation}
{dM \over dl}(l) = {6 \over 5}~{M_0 \over l_{\rm max}}~
\left({l \over l_{\rm max}}\right)^{1 \over 5}~
I_\upsilon\left({l \over l_{\rm max}}\right)
\label{resul}
\end{equation}
where
\begin{equation}
I_{\upsilon}(r) = N_\upsilon~\int_0^{\sqrt{1-r}}~
{dx \over (1-x^2)^{{6 \over 5} - {\upsilon \over 2}}}~~~\ .
\label{Ipfr}
\end{equation}
The predicted angular momentum distributions are shown in Fig. II
for $\upsilon$ = 0, 0.5, 1.0, 1.5, 2.0, 3.0, 5.0, 10., 50. and 100. 
For $\upsilon=0$, ${dM \over dl}(l)$ has a sharp cusp at $l=0$ with 
${dM \over dl}(0) = 3 {M_0 \over l_{\rm max}}$.  However, as soon as
$\upsilon>0$, ${dM \over dl} \propto l^{1 \over 5}$ near $l=0$. For 
$\upsilon$ in the range 0.5 to 2.0, ${dM \over dl}(l)$ is qualitatively 
similar to the angular momentum distributions found by van den Bosch et 
al. in dwarf galaxies.

The average angular momentum is 
\begin{equation}
l_{\rm av} = {1 \over M_0} \int_0^{l_{\rm max}}~dl~l~{dM \over dl}(l) 
= l_{\rm max}~{3 \over 11}~{\sqrt{\pi} \over 2^\upsilon}~
({\upsilon \over 2} + 1)~{\Gamma(\upsilon + 2) \over 
\Gamma({\upsilon \over 2} + 1)~\Gamma({\upsilon \over 2} + {5 \over 2})}~~\ .
\label{avam}
\end{equation}
$l_{\rm max} / l_{\rm av}$, plotted in Fig. III, decreases monotonically 
with $\upsilon$, from 2.75 at $\upsilon=0$ to 1.83 at $\upsilon=\infty$.  
In the range $0.5 < \upsilon < 2.0$, $l_{\rm max} / l_{\rm av}$ ranges 
from 2.56 to 2.29.  So we find that in the axion case 1) the 
$l_{\rm max} / l_{\rm av}$ distribution is sharply peaked, like 
the observed distribution, and 2) it is peaked at roughly the 
same value (2.6) as the observed distribution. 

\subsection{Enhanced caustic rings}

The observational evidence in support of the caustic ring halo model 
is summarized in ref.~\cite{MWhalo}.  A large part of that evidence 
is based on the existence of statistically significant correlations 
between bumps in galactic rotation curves, consistent with the 
assumption that some of the bumps are caused by caustic rings of 
dark matter and that the caustic ring radii obey Eq.~(\ref{crr})
\cite{Kinn,MWcr}.  Additional evidence is provided by the fact that 
the bumps in the high resolution inner rotation curve of the Milky 
Way published in ref.~\cite{Clem} are kinky, i.e. they start with 
an upward kink and end with a downward kink \cite{MWcr}. The kinks 
are explained by the fact that the dark matter density diverges at 
caustic surfaces \cite{sing}.  Yet more evidence is provided by the 
existence of a triangular shape in the IRAS (Infrared Astronomical 
Sattelite) map of the galactic plane in one of the two tangent 
directions to the nearest caustic ring ($n$ = 5).  The position 
of the triangular shape coincides in galactic longitude with the 
position of the rise in the rotation curve associated with that 
caustic ring.  The triangular shape is explained as the imprint 
of the gravitational field of the caustic ring on dust and gas 
in the galactic disk.

There is however a puzzle with the interpretation of the evidence: 
the effects attributed to caustic rings are too large compared to 
theoretical expectation.  Specifically, the bumps in the Milky Way 
rotation curve are on average a factor 5 larger \cite{MWcr} than 
expected in the caustic ring model if the infall is isotropic 
\cite{crdm,sing,MWhalo} and if the bumps are due solely to the 
caustic rings themselves.  The sizes of the bumps are not actually 
predicted precisely by the caustic ring model.  They are given as a 
product of two factors, one of which is predicted by the model.  The 
other factor depends on details that the model (in its present state) 
does not predict and which fluctuate from one caustic ring to the next.
Nonetheless, this second factor is expected to be generally of order
one.  There is no reason why its average should be five.  See refs.
\cite{sing,MWcr} for details.

To account for this discrepancy, ref.~\cite{MWcr} proposed that the 
rises in rotation curves are amplified by gas that is gravitationally
attracted to, and accreted onto, the caustic rings.  The square of the 
velocity dispersion of gas in the galactic disk is sufficiently small 
compared to the gravitational potential ripples caused by caustic rings 
that a large amplification factor is plausible.  However, in this proposal 
it is hard to understand the kinkiness of the bumps in the Milky Way 
rotation curve since the gas distribution would follow the gravitational 
potential of the caustic rings, which is much smoother than the density 
of the caustic rings.

The physics discussed in Section III suggests a simpler and more
compelling explanation, to wit that the infalling axion BEC has a 
big vortex along the galactic symmetry axis and hence that the infall 
is not isotropic.  The caustic rings are enhanced because more dark 
matter falls in near the galactic plane.  If the infall rate is given 
by Eq.~(\ref{infr}), the density of the flow producing the caustic ring 
is increased by the factor $N_\upsilon$ given in Eq.~(\ref{Np}).  For 
$N_\upsilon$ to be of order five, $\upsilon$ must be of order forty.
  
There is a restriction on how large $\upsilon$ can be because a 
caustic ring is partly erased if the infalling dark matter is too 
concentrated near the galactic plane. Using the description in 
ref.~\cite{sing}, one readily finds that caustic rings are formed 
in the flow of particles whose declination $\alpha \equiv {\pi \over 2} 
- \theta$ at their last turnaround is less in magnitude than 
$\alpha_m = {1 \over 2}\sqrt{u \over s} \tau_0 \simeq \sqrt{p \over 2 a}$
where $p$ is the width of the caustic ring.  See ref.~\cite{sing} for 
definitions of $u$, $s$, $\tau_0$.  Some of the caustics may be partly 
erased but not $n=5$ since, as mentioned above, its full cross-section 
appears in IRAS maps of the galactic plane; see 
http://www.phys.ufl.edu/~sikivie/triangle/index.htm.  For that 
ring, $p = 0.018 a$ and hence $\alpha_m = 5.5^\circ$.  Requiring 
$(\cos\alpha_m)^\upsilon > 0.5$ allows $\upsilon$ as large as 150 
and hence an enhancement by the factor $N_\upsilon$ as large as
10.  The rise in the Milky Way rotation curve associated with the 
$n=5$ ring is a factor three larger \cite{MWcr} than expected in 
the isotropic model ($\upsilon = 0$), and therefore consistent 
with the appearance of the full cross-section of the caustic ring 
in the aforementioned IRAS map.  

If the existence of a big vortex is the correct explanation for 
the enhanced effect of caustic rings of dark matter on galactic 
rotation curves, we may derive a lower limit on the fraction $X_a$ 
of dark matter that is axions.  Let $X_W = 1 - X_a$ be the dark 
matter fraction in WIMPs or some other form of ordinary cold dark 
matter.  We need 
\begin{equation}
X_a~N_\upsilon~+~X_W~N_{\upsilon^\prime} \simeq 5
\label{min}
\end{equation}
to account for the average strength of the rises in the Milky 
Way rotation curve.  On the other hand $N_\upsilon \lesssim 10$, 
otherwise the corners of the triangular feature in the IRAS map 
get erased.  Also $\upsilon^\prime \lesssim 5$, otherwise the 
angular momentum distribution of baryonic matter becomes too 
dissimilar to the distributions observed by van den Bosch et 
al.; compare Figs. II and III.  Eq.~(\ref{Np}) implies then 
$N_{\upsilon^\prime} \lesssim 2$.  Combining all this, one 
obtains
\begin{equation}
X_a \gtrsim {3 \over 8}~~~\ ,  
\label{axmin}
\end{equation}
i.e. a lower limit of approximately 37.5\% on the axion dark 
matter fraction.

\section{Summary}

The goal of this paper was to increase our understanding of 
the behaviour of axion BEC dark matter before it falls into 
the gravitational potential well of a galaxy.  In particular 
we wanted to see how axion BEC vortices appear and evolve, 
and whether they have implications for observation.

In Section II, we discussed the properties of rotating many body 
systems in thermal equilibrium.   We showed that the widely used 
self-gravitating isothermal sphere model is an unacceptably poor 
description of galactic halos as soon as angular momentum is 
introduced.  We showed that the vortices that appear in a BEC 
of quasi-collisionless particles, such as an axion BEC, attract 
each other, in contrast to the repulsive behavior of the vortices 
in superfluid $^4$He and dilute gases.  We showed that vortices 
in any BEC appear only as part of the process of rethermalization 
after the BEC has been given angular momentum.  Neither the 
thermalization of a BEC nor the appearance of its vortices 
is described by the Gross-Pitaevskii equation.  That equation 
describes the behavior of the BEC, including the motion of its 
vortices, only after it has formed. 

In Section III, we used the results of Section II to try and 
predict the behavior of axion BEC dark matter before it falls 
into the gravitational potential well of a galaxy.  As angular 
momentum is acquired by the axion BEC through tidal torquing, 
the axions go to a state where all spherical shells rotate 
rigidly about a common axis, with angular velocity proportional 
to $r^{-2}$ where $r$ is the radius of the shell.  The resulting 
angular momentum distribution on the turnaround sphere, Eq.~(\ref{spec}), 
is precisely and in all respects that which accounts for the evidence 
for caustic rings of dark matter.  Because axion BEC vortices are 
attractive, we expect that most join into one big vortex.  The 
radius of this big vortex is smaller than but of order the radius 
of the first caustic ring made by the axion BEC as it falls in 
and out of the galaxy.  We modified the caustic ring model of 
galactic halos to include the presence of the big vortex.  
Whereas the previous version of that model assumed that the 
infall is isotropic, the new version assumes that the infall 
rate is given by Eq.~(\ref{infr}) where $\upsilon$ parametrizes 
the size of the big vortex.

The rate at which baryons and WIMPs reach thermal equilibrium with the 
axion BEC was found to be qualitatively the same as the rate at which 
axions reach thermal equilibrium among themselves.  That thermalization 
rate is larger than the Hubble rate provided the axion dark matter 
fraction is more than of order 3\%.  Because baryons and WIMPs are 
much heavier than axions, the temperature of baryons and WIMPs 
is effectively zero when they are in thermal contact with the 
axions. In that case, baryons and WIMPs acquire the same velocity 
distribution as the axion BEC before falling onto galactic halos, 
and WIMPs produce the same caustic rings as axions do, and at the 
same locations.  We expect the baryons and WIMPs to produce their 
own big vortex although with radius smaller than the radius of the 
big vortex in axions.  The specific angular momentum distribution 
of baryons and WIMPs on the turnaround sphere is the same as for 
axions.  The infall rate is also the same but with a smaller value 
$\upsilon^\prime$ of the parameter $\upsilon$.

In Section V, we compared the specific angular momentum distribution 
predicted for baryons, when the axion dark matter fraction is more than 
of order 3\%, with the specific angular momentum distribution of baryons 
observed in dwarf galaxies.  They are qualitatively similar for
$0.5 \lesssim \upsilon^\prime \lesssim 2$.  Moreover, in this range the 
ratio $l_{\rm max} / l_{\rm av}$ of maximum to average specific angular 
momentum is predicted to be near 2.4.  This is in qualitative agreement 
with the observed distributions since most of the latter have 
$l_{\rm max} / l_{\rm av} \simeq 2.6$.  In contrast, if the dark 
matter is all WIMPs, the specific angular momentum distribution differs 
markedly from the observed distributions; see Fig. I.  Furthermore, 
$l_{\rm max} / l_{\rm av}$ is predicted in the WIMP case to vary from 
galaxy to galaxy, the median value being 4 and the 90\% range from 
2.6 to 8.1.  The ability of axion dark matter to qualitatively explain 
the observed angular momentum of baryons in dwarf galaxies is further 
evidence that at least part of the dark matter is axions.

The appearance of a large vortex in the axion BEC provides a plausible 
solution to a past puzzle, namely that the rises in galactic rotation 
curves attributed to caustic rings of dark matter are typically a factor 
5 larger than expected when the dark matter infall is assumed to be 
isotropic.  The presence of a big vortex implies that more dark matter 
falls in along the galactic plane and hence that the density of the 
flows producing the caustic rings is increased.  If all the dark matter 
is axions, the factor 5 enhancement is accounted for if $\upsilon$ is 
of order 40.  If the dark matter is partly axions and partly WIMPs, with 
the axion fraction more than of order 3\%, axions and WIMPs co-produce
the caustic rings.  If one requires $\upsilon^\prime < 5$ to have an 
acceptable fit between the predicted and observed specific angular 
momentum distribution of  baryons in dwarf galaxies, the caustic 
ring enhancement of a factor of five can only be accounted for 
if at least 37\% of the dark matter is axions.

\begin{acknowledgments}

We thank Leslie Rosenberg, Tom Quinn, Tanja Rindler-Daller, Paul Shapiro,
Frank van den Bosch and Mark Srednicki for stimulating discussions.  This 
work was supported in part by the U.S. Department of Energy under grant 
DE-FG02-97ER41209 at the University of Florida, the National Science 
Foundation under Grant No. PHYS-1066293 at the Aspen Center for Physics, 
and the National Science Foundation under grant PHY11-25915 at the Kavli 
Institute for Theoretical Physics at UC Santa Barbara.

\end{acknowledgments}


\newpage



\newpage

\begin{figure}
\begin{center}
\includegraphics[angle=360, width=160mm]{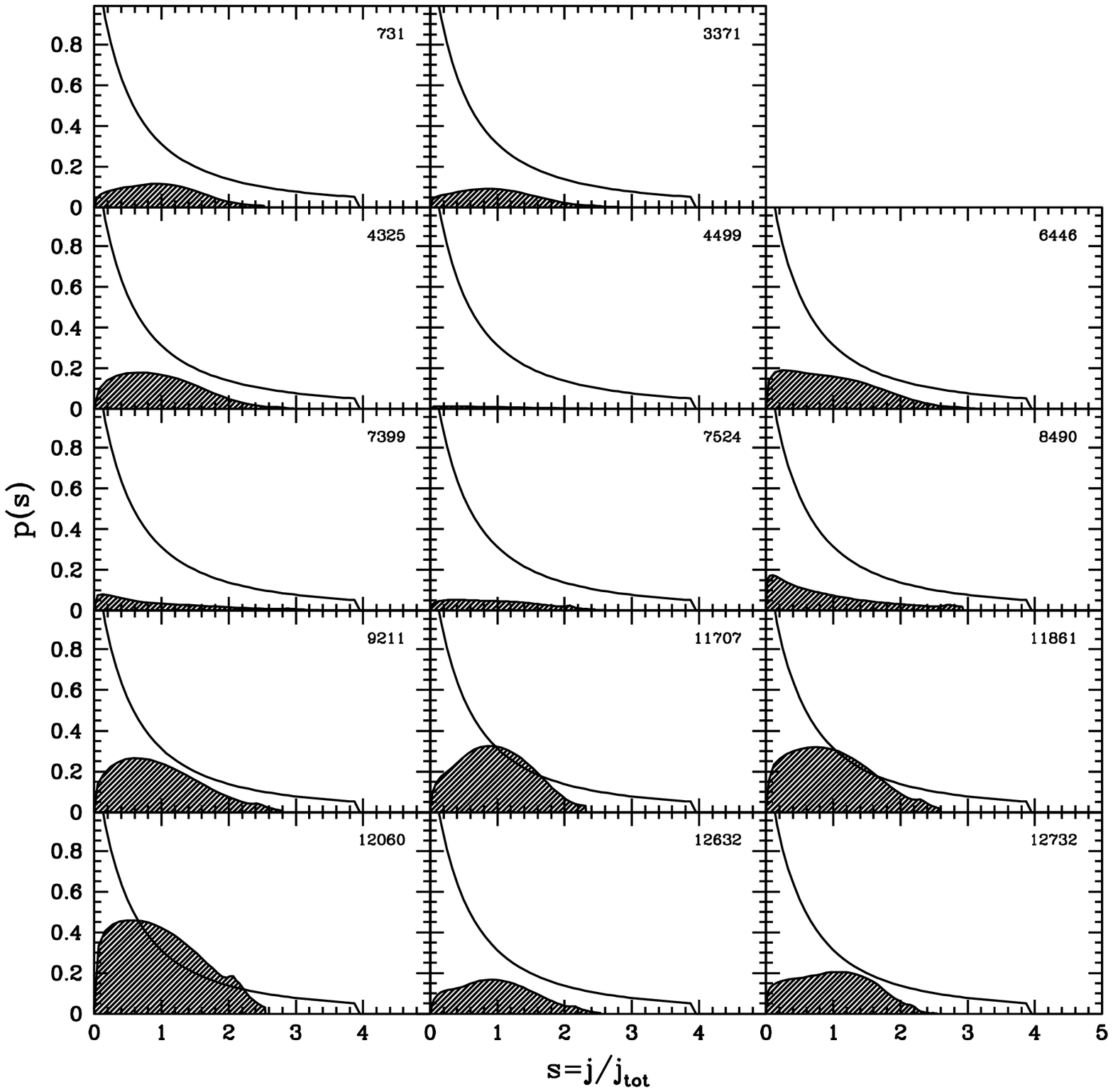}
\vspace{0.2in}
\caption{Reproduction of Fig. 4 in the article {\it The angular 
momentum content of dwarf galaxies: new challenges for the theory 
of galaxy formation} by F.C. van den Bosch et al. \cite{Bosch}.  The 
shaded areas indicate the specific angular momentum distributions of 
baryons in fourteen dwarf galaxies.  In terms of the quantities defined in 
the text, $s = {l \over l_{\rm av}}$ and $p(s) \propto {d M \over dl}(l)$.  
The solid curve is the distribution, Eq.~(\ref{numdis}), predicted
by numerical simulations of galaxy formation with ordinary cold 
dark matter, for $\mu$ = 1.25.}
\end{center}
\label{vdBB4}
\end{figure}

\newpage
\vspace{-3in}
\begin{figure} 
\begin{center}
\includegraphics[angle=360, height=200mm]{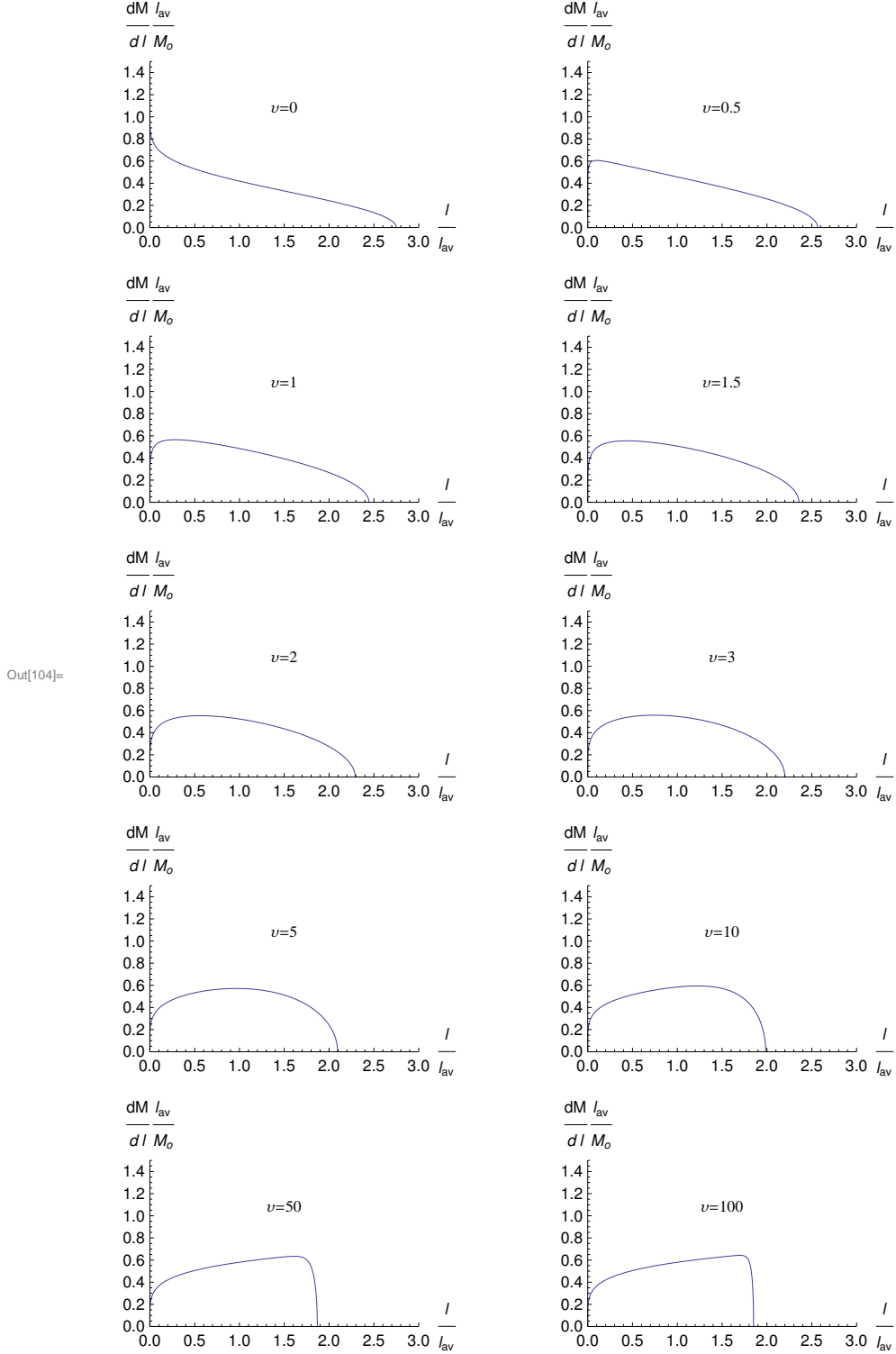}
\vspace{0.3in}
\caption{Specific angular momentum distributions if the 
dark matter is axions, for various values of the parameter
$\upsilon$.}
\end{center}  
\label{amd}
\end{figure}

\newpage

\begin{figure} 
\begin{center}
\includegraphics[angle=360, width=150mm]{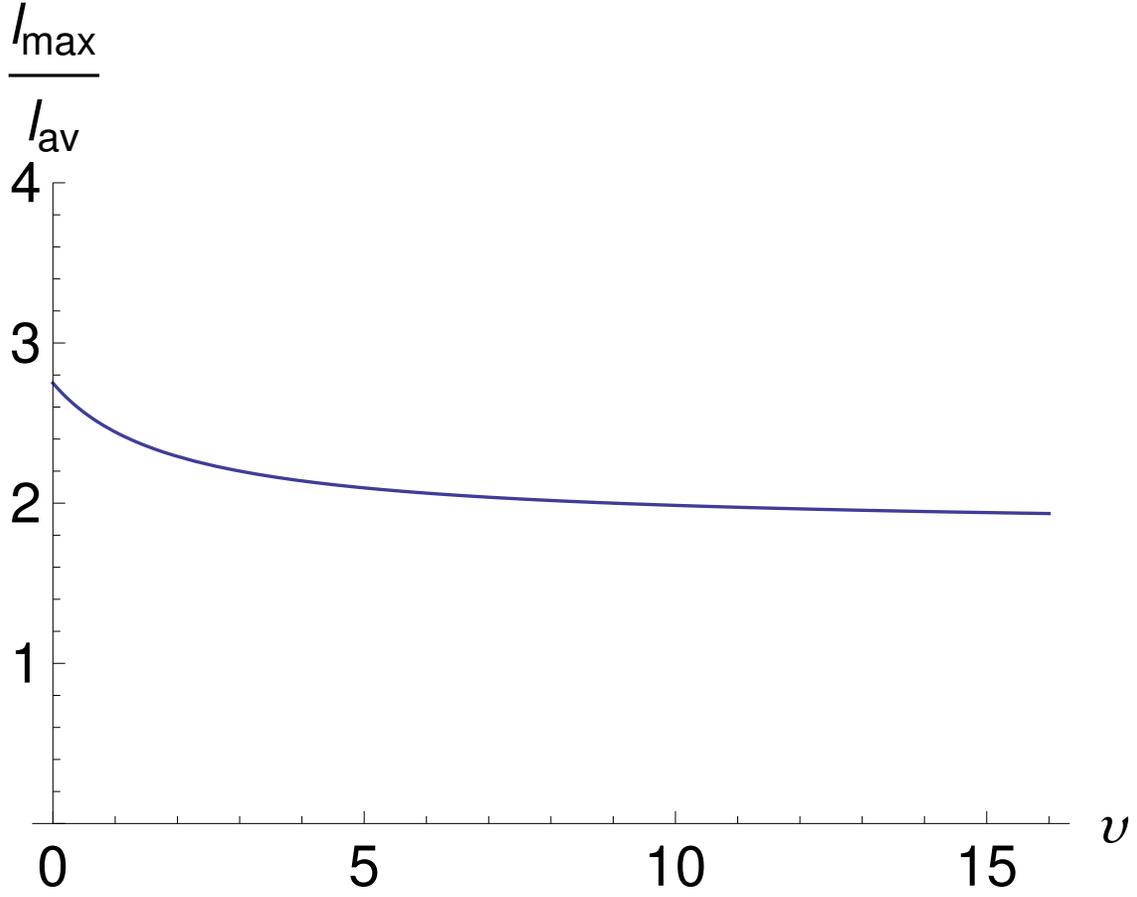}
\vspace{0.3in}
\caption{The ratio of maximum to average angular 
momentum if the dark matter is axions, as a function
of the parameter $\upsilon$.  For the fourteen dwarf 
galaxies observed by van den Bosch et al. this ratio
is narrowly peaked near 2.6.  In numerical simulations
of galaxy formation with ordinary cold dark matter, this
ratio is predicted to vary from galaxy to galaxy over the 
range 2.6 to 8.1 (90 \%CL), with median value 4.0}
\end{center}  
\label{lratio}
\end{figure}


\end{document}